\documentclass[12pt]{article}
\usepackage[utf8]{inputenc}

\usepackage{times}

\usepackage{graphicx}
\usepackage{amsmath}
\usepackage{appendix}
\usepackage{amsfonts}
\usepackage[normalem]{ulem}
\usepackage{siunitx}
\usepackage{xcolor}
\usepackage{float}

\newcommand{\be}{\begin{equation}}
\newcommand{\ee}{\end{equation}}
\newcommand{\bea}{\begin{equation} \begin{aligned}}
\newcommand{\eea}{\end{aligned} \end{equation} }
\newcommand{\bpm}{\begin{pmatrix}}
\newcommand{\epm}{\end{pmatrix}}
\newcommand{\mbf}[1]{\mathbf{#1}}

\usepackage[
 autocite=superscript,
 style=nature,
 sortcites= true,
 backend=biber,
 isbn=false,
 date=year,
 doi=false,
 url=false,
 eprint=false,
]{biblatex}

\AtEveryBibitem{\clearfield{issue}}

\usepackage{lineno}

\title{Bulk and Edge Properties of Twisted Double-Bilayer Graphene}

\author
{Yimeng Wang$^{1}$, Jonah Herzog-Arbeitman$^{2}$, G. William Burg$^{1}$, Jihang Zhu$^{3}$,\\ 
Kenji Watanabe$^{4}$, Takashi Taniguchi$^{5}$, Allan H. MacDonald$^{3}$,\\ B. Andrei Bernevig$^{2}$, Emanuel Tutuc$^{1,\ast}$ \\
\normalsize{$^{1}$Microelectronics Research Center, Department of Electrical and Computer Engineering,}\\
\normalsize{The University of Texas at Austin, Austin, TX 78758, USA}\\
\normalsize{$^{2}$Department of Physics, Princeton University, Princeton, New Jersey 08544, USA}\\ 
\normalsize{$^{3}$Department of Physics, University of Texas at Austin, Austin, TX 78712}\\
\normalsize{$^{4}$Research Center for Functional Materials, National Institute of Materials Science,}\\
\normalsize{$^{5}$International Center for Materials Nanoarchitectonics, National Institute of Materials Science,}\\ 
\normalsize{1-1 Namiki Tsukuba, Ibaraki 305-0044, Japan}\\
\normalsize{$^\ast$ Correspondence author:etutuc@mail.utexas.edu}
}

\date{}

\begin{document}

\baselineskip24pt

\maketitle

\textbf{The emergence of controlled, two-dimensional moir\'e materials \autocite{cao2018, cao2018a, burg2019, shen2020, cao2020, liu2020} has uncovered a new platform for investigating topological physics \autocite{wu2019,chebrolu2019,koshino2019}. Twisted double bilayer graphene (TDBG) has been predicted to host a topologically non-trivial gapped phase with Chern number equal to two at charge neutrality, when half the flat bands are filled\autocite{chebrolu2019,koshino2019}. However, it can be difficult to diagnose topological states using a single measurement because it is ideal to probe the bulk and edge properties at the same time. Here, we report a combination of chemical potential measurements, transport measurements, and theoretical calculations that show that twisted double bilayer graphene can host metallic edge transport while simultaneously being insulating in the bulk. A Landauer-B\"{u}ttiker analysis of measurements on multi-terminal samples allows us to quantitatively assess edge state scattering. We interpret these results as signatures of the predicted topological phase at charge neutrality.}

The tunability of moir\'e materials \autocite{cao2018, cao2018a, burg2019, shen2020, cao2020, liu2020} by angle and carrier density powers the realization of novel topological phases \autocite{wu2019,chebrolu2019,koshino2019}, such as correlated Chern insulators that break time-reversal-symmetry \autocite{sharpe2019,serlin2020,2020arXiv200703735W,2020arXiv200713390D,park2020,2020arXiv200703810N,2020arXiv200811746C}.
The long moir\'e period justifies low-energy effective models that neglect \autocite{2011PNAS..10812233B} weak scattering between distant regions of momentum space known as valleys.  The application of a transverse electric field causes bulk TDBG to open a gap at charge neutrality to a topological state analogous to the quantum spin Hall (QSH) state characterized by a non-zero valley Chern number $C_V = 2$ \autocite{chebrolu2019, koshino2019}, indicating two pairs of counter-propagating edge states per spin. However, disruption of the  moir\'e pattern on the edge is expected to break the valley symmetry, leaving the states susceptible to scattering and their fate uncertain.

A simultaneous investigation of bulk and edge properties is necessary to reveal the band topology of an electronic system. We employ a sample design shown in Fig. 1a that allows simultaneous electron transport and chemical potential measurements. Our samples consist of a double layer where one layer is a controlled moir\'e material TDBG, with a twist angle ($\theta$) range of $0.97^\circ - 1.60^\circ$. The second layer is a graphene back-gate (GrBG), consisting of monolayer or bilayer graphene with terminals for resistance measurements. The TDBG and GrBG layers are separated by a hexagonal boron-nitride (hBN) dielectric. The GrBG acts not only as the back-gate, but as a Kelvin probe of the TDBG chemical potential \autocite{kim2012, lee2014}. The double layer is encapsulated in hBN, with an added graphite top-gate, and placed on a SiO$_2$/Si substrate, which serves as an additional gate. This sample architecture allows access to the chemical potential of the TDBG in a wide range of carrier density ($n$) and transverse electric field ($E$). Similar double-layers have been used to probe chemical potentials and thermodynamic gaps in bilayer graphene\autocite{lee2014}, and twisted bilayer graphene\autocite{park2020}, albeit without control of the transverse electric field. An optical micrograph of the sample is illustrated in Fig. 1b.

The longitudinal resistance ($R_\mathrm{xx}$) of the TDBG is measured as a function of top-gate bias ($V_\mathrm{TG}$) and graphene back-gate bias ($V_\mathrm{BG}$) to determine the twist angle, and characterize the sample quality. The $V_\mathrm{TG}$ and $V_\mathrm{BG}$ values tune $n$ and $E$ independently according to $n=(C_\mathrm{TG}V_\mathrm{TG}+C_\mathrm{BG}V_\mathrm{BG})/e$, and $E=(C_\mathrm{TG}V_\mathrm{TG}-C_\mathrm{BG}V_\mathrm{BG})/2\epsilon_\mathrm{0}$, where $e$ is the electron charge, $\epsilon_\mathrm{0}$ is the vacuum permittivity, and $C_\mathrm{TG}$ and $C_\mathrm{BG}$ are the capacitances per unit area of the top- and back-gate, respectively. The values of $C_\mathrm{TG}$ and $C_\mathrm{BG}$ can be first determined from the dielectric thickness, and confirmed with magnetotransport measurements (see Supplementary Information section A). Figure 1c shows the contour plot of $R_\mathrm{xx}$ as a function of $n$ and $E$, which exhibits resistance maxima at densities commensurate with $n_\mathrm{s}=2.2\times 10^{12}$ cm$^{-2}$ associated with filling of one moir\'e band with 4-fold spin-valley degeneracy. Resistance maxima are observed at $n=0, \pm n_\mathrm{s}, \pm 3n_\mathrm{s}$, consistent with single-particle band structure calculations, as well as correlated insulators at $n=\pm n_\mathrm{s}/2$\autocite{burg2019, shen2020, liu2020, cao2020}. The $R_\mathrm{xx}$ maxima at $n=\pm 2n_\mathrm{s}$ are a surprise because no gap between the second and third moir\'e bands is predicted in single-particle band calculations. We tentatively attribute the $R_\mathrm{xx}$ maxima at $n=\pm 2n_\mathrm{s}$ to a gap opening driven by electron-electron interactions. The twist angle ($\theta$) can be extracted using the equation $n_\mathrm{s} =\frac{8}{\sqrt{3}}\left(2 a^{-1} \sin (\theta/2) \right)^2$, where $a=\SI{2.46}{\angstrom}$ is the graphene lattice constant. Figure 1d shows a line cut of $R_\mathrm{xx}$ vs. $n$ measured along the dashed line in Fig. 1c. Sharp $R_\mathrm{xx}$ peaks at integer $n/n_\mathrm{s}=0, \pm 1, -2, +3$, and fractional $n/n_\mathrm{s}=+1/2$ further illustrate the observations  in Fig. 1c, and highlight the high TDBG sample quality.

We focus primarily on the TDBG properties at charge neutrality where one of the two flat bands, per valley per spin is filled. The bulk topology\autocite{burg2020} can be determined from the Bistritzer-MacDonald Hamiltonian\autocite{2011PNAS..10812233B}. Figure 1e 
shows how the twisting of the two graphene Brillouin zones forms the moir\'e Brillouin zone for the $K$ valley. 
In Fig. 1f, we show the calculated band structure at an applied interlayer potential $V=\SI{15}{meV}$ using the parameters in Ref. \autocite{burg2020}, and highlight the highest valence band at charge neutrality, labelled $-1$. This band carries a Chern number $+2$, and because all lower bands carry zero Chern number, the total Chern number of the occupied bands is $C_K =2$. The $K'$ valley is related to $K$ by time-reversal and must have opposite Chern number. Therefore the phase is characterized by the valley Chern number $C_V = (C_K - C_{K'})/2 = 2$ at charge neutrality. 

By tuning the GrBG doping ($n_\mathrm{BG}$) with a substrate bias ($V_\mathrm{s}$), we are able to directly probe the TDBG chemical potential ($\mu$) as a function of both $n$ and $E$. An analysis of the band alignment in the heterostructure (see Supplementary Information section B) shows that when the GrBG is charge neutral ($n_\mathrm{BG}=0$), the TDBG chemical potential satisfies:  
\begin{equation}
    \mu=eV_\mathrm{BG} \left( 1+\frac{C_\mathrm{s}}{C_\mathrm{BG}}\right)-eV_\mathrm{s} \frac{C_\mathrm{s}}{C_\mathrm{BG}},
\end{equation}
where $C_\mathrm{s}$ is the capacitance per unit area between the substrate and the GrBG. To determine the charge neutrality gap at different $E$-fields, we utilize Eq. (1) and Fig. 2a data, which shows a contour plot of $R_\mathrm{xx}$ vs. $V_\mathrm{TG}$ and $V_\mathrm{BG}$ for a TDBG sample with $\theta=0.97^\circ$. Along the black dashed diagonal the TDBG density is $n=0$,
while the $E$-field varies. By mapping the longitudinal resistance of the GrBG ($R_\mathrm{BG}$) vs. $V_\mathrm{TG}$ and $V_\mathrm{BG}$ and tracking the GrBG charge neutrality, $\mu$ can be extracted according to Eq. (1). Figure 2b-d show three contour plots of $R_\mathrm{BG}$ vs. $V_\mathrm{TG}$ and $V_\mathrm{BG}$ in the vicinity of $n_\mathrm{BG}=0$, measured at different $V_\mathrm{s}$ values. The corresponding $V_\mathrm{TG}$ and $V_\mathrm{BG}$ ranges used in Figs. 2b-d are marked by rectangles in Fig. 2a. By changing the $V_\mathrm{s}$ value at which the $R_\mathrm{BG}$ vs. $V_\mathrm{TG}$ and $V_\mathrm{BG}$ data are acquired, the $E$-field value at the intersection point of the $n=0$ and $n_\mathrm{BG}=0$ lines can be tuned accordingly. Indeed, the TDBG charge neutrality gaps are measured at three $E$ values in Figs. 2b-d -- $\SI{0.15}{\V/\nm}$ (panel b), $\SI{0.37}{\V/\nm}$ (panel c), and $\SI{0.29}{\V/\nm}$ (panel d). The black dashed lines in Figs. 2b-d 
illustrate the evolution of GrBG charge neutrality with $V_\mathrm{TG}$ and $V_\mathrm{BG}$, 
which can be readily converted into a $\mu$ vs. $n$ dependence using Eq. (1). 
The clear step in the GrBG charge neutrality line observed as it crosses the TDBG $n=0$ line shown in Figs. 2c and 2d 
reveals a thermodynamic gap at the TDBG charge neutrality.  
In contrast to Figs. 2c and 2d data, in Fig. 2b the $n_\mathrm{BG}=0$ line is flat at $n=0$, indicating the absence of a gap at low $E$-field. Due to our emphasis on TDBG charge neutrality in this work, we show $R_\mathrm{BG}$ vs. $V_\mathrm{TG}$ and $V_\mathrm{BG}$ only in the vicinity of double neutrality $n= n_\mathrm{BG}=0$ in Figs. 2b and 2c. However, the $\mu$ values can also be probed away from $n=0$. Indeed, Fig. 2d shows a contour plot of $R_\mathrm{BG}$ vs. $V_\mathrm{TG}$ and $V_\mathrm{BG}$ that captures a wider TDBG density range, and displays features that indicate gaps at other moir\'e filling factors. 

Figure 2e summarizes the $\mu$ vs. $n$ dependence near $n=0$, at varied $E$ values. The $\mu$ vs. $n$ step at $n=0$ marks the opening of a gap ($\Delta_\mathrm{0}$) which increases with the applied $E$-field. The evolution of $\Delta_\mathrm{0}$ vs. $E$ for two samples with twist angles $\theta=0.97^\circ$ and $\theta=1.60^\circ$ is shown in Fig. 3a. For comparison Fig. 3a data include the gap at neutrality measured in Bernal stacked bilayer graphene as a function of $E$-field\autocite{lee2014}. In both samples the $\Delta_\mathrm{0}$ values exceed 100 K at large $E$-fields, and are as high as 300 K in the $\theta=0.97^\circ$ TDBG.

Figure 3b shows the longitudinal resistivity ($\rho_\mathrm{xx}$) vs. $E$ measured at $n=0$ for three TDBG samples with $\theta$ values between $0.97^\circ$ and $1.60^\circ$. The $E$-field range where $\Delta_\mathrm{0}$ opens in the $\theta=0.97^\circ$ sample is highlighted. The $\rho_\mathrm{xx}$ values remain low in the $E$-field range where $\Delta_\mathrm{0}$ is negligible, and show a sharp increase concomitant with the gap opening. Surprisingly, the TDBG $\rho_\mathrm{xx}$ values then decrease with increasing $E$, and remain well within the expected range of a metallic electron system ($<h/e^2$, where $h$ is Planck's constant) despite the gap opening for all three samples. We note that similar low $R_\mathrm{xx}$ values at $n=0$ and at high $E$-fields can be seen in previous studies \autocite{sinha2020, he2021}. Figure 3b also shows $\rho_\mathrm{xx}$ vs. $E$ at $n=0$ data for a Bernal stacked bilayer graphene \autocite{lee2014}. While both the TDBG and Bernal stacked bilayer graphene have a $E$-field dependent gap at neutrality, the $\rho_\mathrm{xx}$ vs. $E$ dependence of the Bernal stacked graphene is markedly different, quickly reaching large values with increasing $E$, as expected for a prototypical band insulator. In Fig. 3c we show the ratio of the nonlocal resistance ($R_\mathrm{NL}$) to $R_\mathrm{xx}$ as a function of $n$ and $E$ measured in the $\theta=0.97^\circ$ TDBG. The measurement configuration schematics for $R_\mathrm{NL}$ and $R_\mathrm{xx}$ are shown in the Fig. 3c insets. These data show a large $R_\mathrm{NL}/R_\mathrm{xx}$ ratio at charge neutrality and at high $E$-fields when $\Delta_\mathrm{0}$ opens, which signals a distinct change in the current pattern. It is noteworthy that both $R_\mathrm{NL}$ and $R_\mathrm{xx}$ values remain well below $h/e^2$ at charge neutrality, and the $R_\mathrm{NL}/R_\mathrm{xx}$ ratio remains small elsewhere in Fig. 3c contour plot. The sharp increase in $R_\mathrm{NL}/R_\mathrm{xx}$ ratio at charge neutrality combined with the low $\rho_\mathrm{xx}$ values when $\Delta_0$ opens suggests the presence of edge transport when the TDBG is gapped.

The presence of edge transport may not necessarily stem from non-trivial bulk topology. Indeed, electrostatic edge states due to finite sample width can emerge in small-bandgap semiconductors \autocite{nichele2016}. To further test the origin of the edge transport in TDBG, it is important to examine a similar twisted system, but with a topologically trivial gap. To this end, in Fig. 3b we include $\rho_\mathrm{xx}$ vs. $E$ measured at $n=0$ for a TDBG with $\theta=181.10^\circ$, i.e. twisted at a small angle with respect to 180$^\circ$. Additional data measured in a TDBG with $\theta=181.9^\circ$ can be found in Supplementary Information section C. The TDBGs with $\theta \approx 180^\circ$ have a similar band structure as TDBGs with $\theta \approx 0^\circ$, and open a gap at $n=0$ at finite $E$-fields (see Supplementary Information section D). The magnetotransport properties of the $\theta=181.10^\circ$ TDBG (Fig. S2) reveal a rich Hofstadter butterfly comparable to samples with $\theta \approx 0^\circ$ (Fig. S1), indicating a similarly high quality sample. However, thanks to the C$_\mathrm{2y}$ symmetry in TDBGs with $\theta \approx 180^\circ$ the gap at neutrality is topologically trivial with $C_V=0$\autocite{koshino2019}. Figure 3d compares $\rho_\mathrm{xx}$ vs. $E$ at $n=0$ at different temperatures measured in the $\theta=181.10^\circ$ TDBG (left panel) with $\rho_\mathrm{xx}$ vs. $E$ at $n=0$ for the three TDBGs twisted with respect to $0^\circ$ (right panel). The contrast between the two sets of samples is noteworthy --  $\rho_\mathrm{xx}$ quickly reaches values significantly larger than $h/e^2$, with an insulating temperature dependence in the $\theta = 181.10^\circ$ TDBG, while the TDBGs with $\theta \approx 0^\circ$ show $\rho_\mathrm{xx}$ lower than $h/e^2$ when the gap opens at neutrality. The contrast between the $\theta \approx 180^\circ$ TDBG and the $\theta \approx 0^\circ$ TDBGs suggests non-trivial edge transport in the TDBGs with $\theta \approx 0^\circ$, which we associate with the emergence of a topological valley Chern insulator.
   
To better describe the TDBG band structure and topology, we compute a single-particle phase diagram for TDBG\autocite{burg2020} for $\theta = 0.8^\circ - 1.8^\circ$ and varying $V$, controlled by the applied $E$-field. We extract the indirect gap at charge neutrality $\Delta_0$ from the band structure as depicted in Fig. 1f, and compute the Chern numbers of the occupied bands. Figure 4a shows a contour plot of $\Delta_0$ vs. $\theta$ and $V$, which reveals a gap closing and reopening (blue dashed line) for all angles at small $V$. Beyond the gap closing, corresponding to larger $E$-fields, we find a gapped topological phase with $C_V = 2$. The results are consistent with the experimental data in our samples. Interestingly, Fig. 4a data predicts a trivial insulator for $\theta > 1.1^\circ$ with a gap of approximately $\SI{2} {meV}$ at $E=0$ which may account for the $\rho_\mathrm{xx}$ decrease at small $E$-fields for the $\theta=1.60^\circ$ TDBG in Fig. 3b, and observed for TDBGs with $\theta = 1.55-1.9^{\circ}$ in Ref. \autocite{chu2021}. 

 Figures 4b and 4c show the phase diagrams at $n/n_\text{s} = \pm 1$, dominated by a gapless phase and a trivial insulator. However, at $\theta \approx 0.9^\circ$ and small $E$-fields, a gapped state with $C_V = +1$ (shaded area) appears for both $n/n_\text{s} = \pm 1$. In this state at $n/n_\text{s} = +1$ in valley $K$, the upper flat band has Chern number $-1$ which only partially cancels the Chern number $+2$ carried by all valence bands. Similarly, at $n/n_\text{s} = -1$, the dispersive bands below the flat bands carry Chern number $1$, and the lower flat band carrying Chern number $1$ is unoccupied. It is noteworthy that in addition to the large $R_{\mathrm{NL}}/R_{\mathrm{xx}}$ observed at $n=0$, Fig. 3c data shows satellites of finite $R_{\text{NL}}/R_{\text{xx}}$ at $n/n_\text{s} = 1$, a possible signature of the topological state at this filling factor. There is also a $R_{\mathrm{NL}}/R_{\mathrm{xx}}$ peak at $n/n_\text{s} = -1$, but with a weaker nonlocal resistance, indicating some degree of particle-hole symmetry breaking. This finding is consistent with the shaded region of Fig. 4c data which is interrupted by a metallic phase near $\theta =0.97^\circ$. 

To quantitatively probe the edge state transport (Fig. 4d) we employ the Landauer-B\"{u}ttiker (LB) formalism \autocite{PhysRevLett.57.1761,PhysRevB.38.9375,2007Sci...318..766K}. The $\theta=1.60^{\circ}$ TDBG has nine terminals $i = 1,\dots,9$ which enables a variety four-terminal resistance measurements relating the currents $I_i$ and voltages $V_i$ (Fig. 4e). The LB equation, normalized by $2C_V$ to include spin degeneracy, reads
\begin{equation}
\label{eq:LB}
I_i = 2C_V \frac{e^2}{h} \sum_{j=1}^N (T_{ji} V_i - T_{ij} V_j) \ . \\
\end{equation}
Here $T_{ij}$ is the transmission probability from $j \to i$, $N=9$ in our sample. Because the disruption of the moir\'e pattern on the edge and the sample geometry the transmission matrix $T$ is not expected to take a simple form. 
We derive an exact inverse relating resistance measurements to the entries of $T$, extending prior results for a four-terminal sample\autocite{5390001} (see Supplementary Information section E). Using the matrix $T(E)$ we then extract the order parameter 
\begin{equation}
\label{eq:sigma}
\sigma(E) = \frac{2 C_V e^2}{18h}\sum_{i\neq j}^9 T_{ij}(E)
\end{equation}
where $\sigma$, the total edge conductance, is equal to $4 e^2/h$ when edge states do not back-scatter (see Supplementary Information section E.6). Figure 4f shows $\sigma$ vs. $E$ and reveals a divergence in the order parameter at the topological phase transition at $E = \SI{0.33} {V/nm}$. For larger $E$ in the topological phase, $\sigma$ decays quickly to a constant finite value. To further illustrate the edge transport, Fig. 4f shows $||\Delta T(E)||$, the average deviation of $T(E)$ from $T(\SI{.7} {V/nm})$ defined by the Frobenius norm (see Supplementary Information section E.6). A small value of $||\Delta T(E)||$ means that $T(E)$ is close to $T(\SI{0.7} {V/nm})$ in every entry. Once the gap is open, we find that $||\Delta T(E)|| \approx 0$, indicating that the whole matrix $T(E)$ and the edge states it describes are independent of $E$, suggestive of topological effects. 

Lastly, we comment on the contrast with Bernal stacked bilayer graphene, a material with large Berry curvature at the Brillouin zone corners, and theoretically expected to possess edge states for specific terminations \autocite{2016PhR...648....1R,PhysRevB.92.115437}. In TDBG, we find the transverse edge state localization length $\xi \approx \SI{100}{nm}$, whereas $\xi \approx \SI{1}{nm}$ in Bernal graphene (Supplementary Information section F), making the edge states much more sensitive to edge disorder. This implies that edge states scattering in TDBG is significantly reduced thanks to the long moir\'e period.

In summary, simultaneous thermodynamic and transport properties can provide unique insights into band topology effects in moir\'e materials, particularly for states that do not break time-reversal symmetry. The data suggests the emergence of a tunable topological insulator in TDBG, as a consequence of non-zero valley Chern numbers of the moir\'e bands.

\section*{Acknowledgements}

We thank Zhi-Da Song and Biao Lian for helpful discussions. 
The work at The University of Texas was supported by the National Science Foundation Grants MRSEC DMR-1720595, EECS-1610008, EECS-2122476, Army Research Office under Grant No. W911NF-17-1-0312, and the Welch Foundation grant F-2018-20190330. Work was partly done at the Texas Nanofabrication Facility supported by NSF Grant No. NNCI-1542159. B.A.B. was
supported by the DOE Grant No. DE-SC0016239, the
Schmidt Fund for Innovative Research, Simons Investigator Grant No. 404513, the Packard Foundation, the
Gordon and Betty Moore Foundation through Grant No.
GBMF8685 towards the Princeton theory program, and
a Guggenheim Fellowship from the John Simon Guggenheim Memorial Foundation. Further support was provided by the NSF-EAGER No. DMR 1643312, NSFMRSEC No. DMR-1420541 and DMR-2011750, ONR
No. N00014-20-1-2303, BSF Israel US foundation No. 2018226,
and the Princeton Global Network Funds. J.H.A. acknowledges support from a Marshall Scholarship funded by the Marshall Aid Commemoration Commission. A.H.M. acknowledges support from Welch Foundation grant F1473. K.W. and T.T. acknowledge support from the Elemental Strategy Initiative conducted by the MEXT, Japan (Grant Number JPMXP0112101001), JSPS KAKENHI (Grant Numbers JP19H05790 and JP20H00354).

\printbibliography
\clearpage

\begin{figure}[H]
\begin{center}
\includegraphics[width=6.3in]{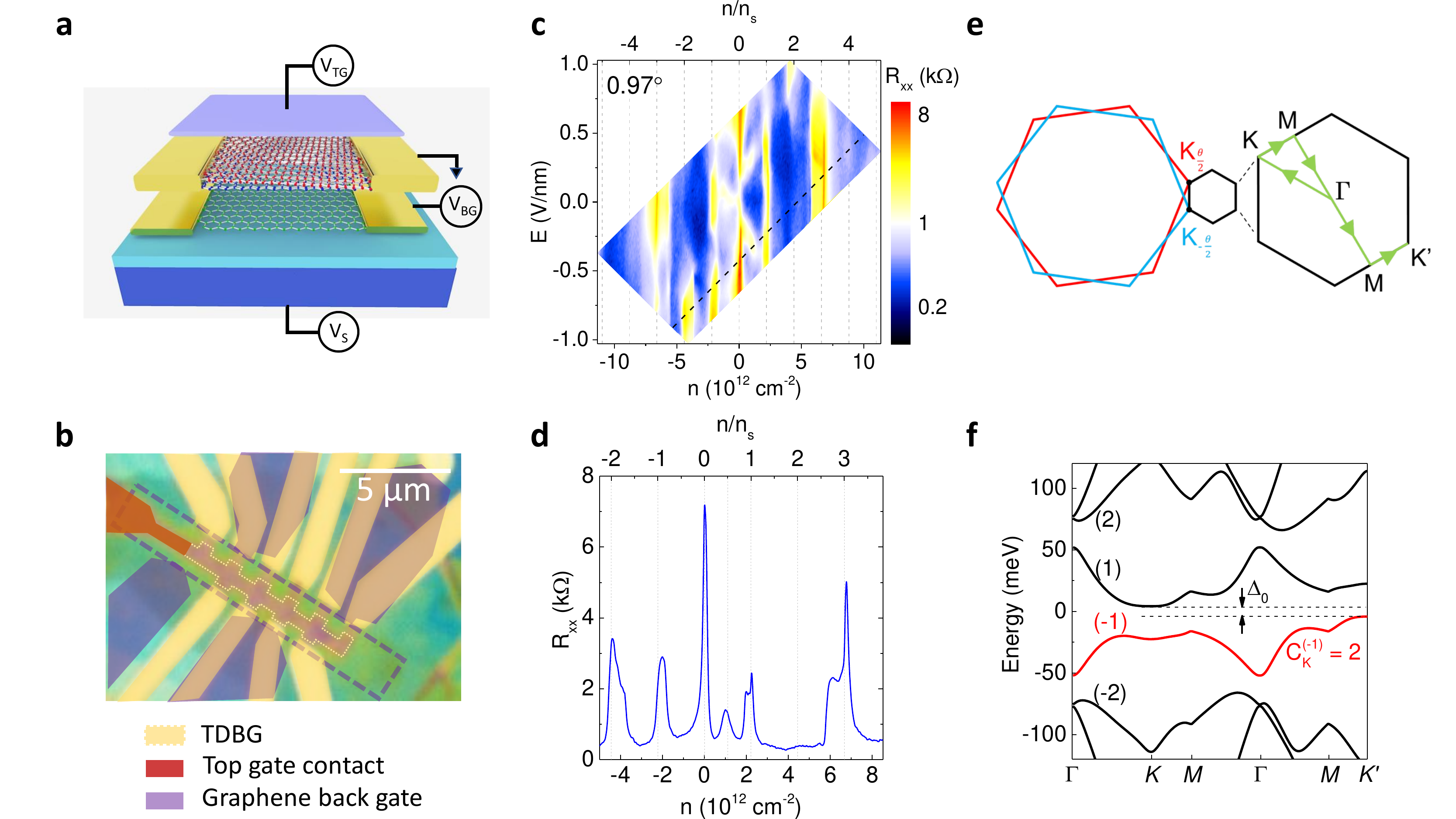}
\end{center}
\caption{{\bf Sample layout, electrical transport, and band structure. a}, Schematic of the TDBG sample structure, with graphite top-gate and monolayer or bilayer graphene back-gate. The substrate is separated from the dual-gated TDBG by a dielectric layer. The top-gate, back-gate, and substrate are individually biased. {\bf b}, Optical micrograph of a TDBG sample. The active area, top gap, and back-gate are marked. {\bf c}, $R_\mathrm{xx}$ vs. $n$ and $E$, measured in a TDBG samples with $\theta=0.97^\circ$ at $T=\SI{1.5}{K}$. {\bf d}, $R_\mathrm{xx}$ vs. $n$ measured  along the dashed line in {\bf c}. The top axis shows the moir\'e band filling factor $n/n_\mathrm{s}$ in panels {\bf c} and {\bf d}. The data show $R_\mathrm{xx}$ maxima at integer and fractional filling factors. {\bf e}, Schematic of the moir\'e Brillouin zone. {\bf f}, Calculated moir\'e band structure of TDBG with $\theta=1.60^\circ$ in the $K$ valley at $V=\SI{15}{meV}$, with the charge-neutrality gap $\Delta_0$ shown. The Chern number of the first valence band in the $K$ valley is 2.}
\label{fig1}
\end{figure}

\begin{figure}[H]
\begin{center}
\includegraphics[width=6.3in]{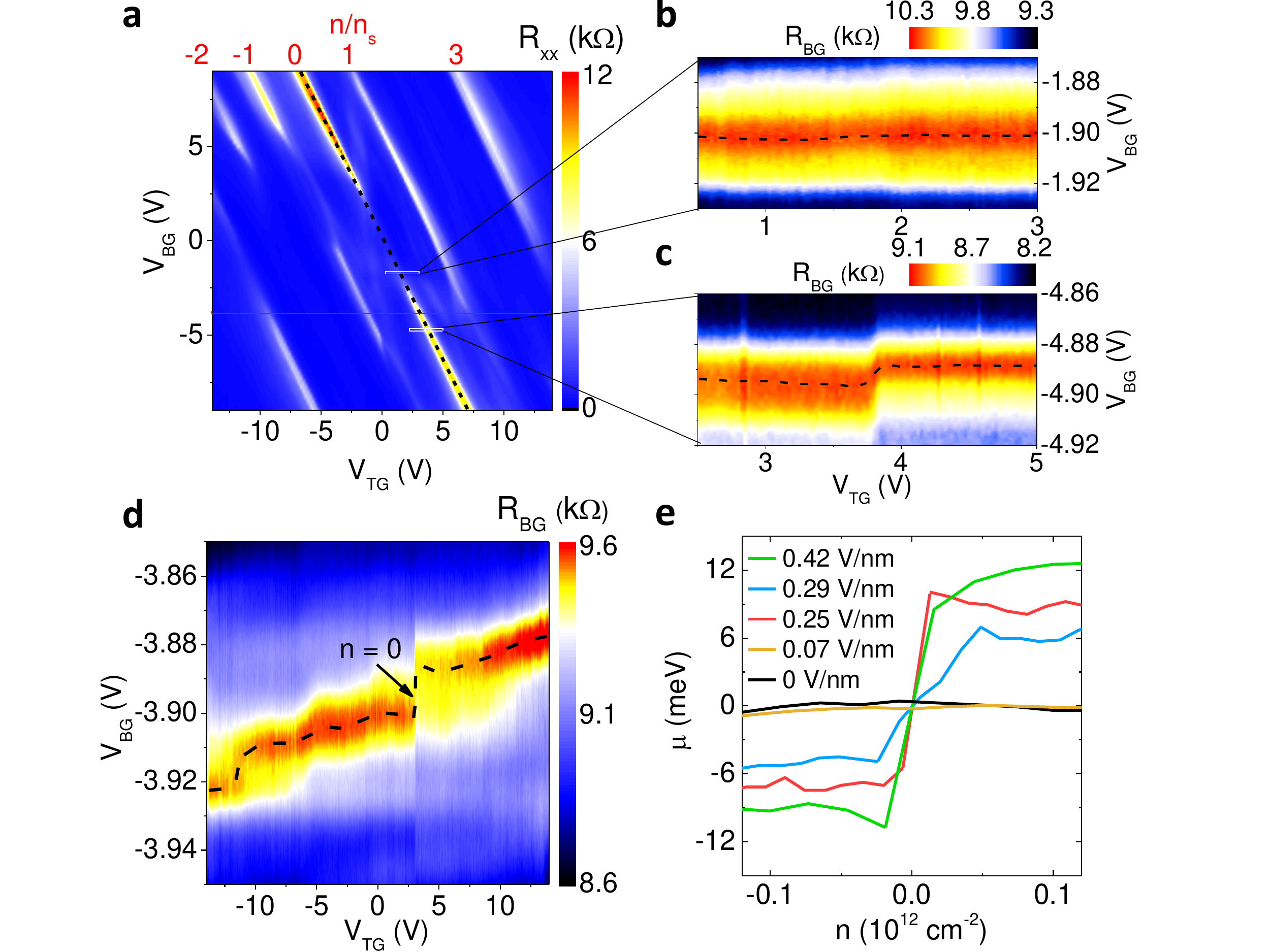}
\end{center}
\caption{{\bf Chemical potential measurements. a}, Contour plot of $R_\mathrm{xx}$ vs. $V_\mathrm{TG}$ and $V_\mathrm{BG}$ measured in a TDBG sample with $\theta=0.97^\circ$. The moir\'e band filling factors corresponding to $R_\mathrm{xx}$ maxima are labeled in red. The black dashed line marks the TDBG charge neutrality. {\bf b}-{\bf d}, $R_\mathrm{BG}$ vs. $V_\mathrm{TG}$ and $V_\mathrm{BG}$. The black dashed lines mark the $R_\mathrm{BG}$ maxima indicating to the GrBG charge neutrality loci ($n_\mathrm{BG}=0$). The corresponding $V_\mathrm{TG}$ and $V_\mathrm{BG}$ ranges used in panels {\bf b-c} and {\bf d} are marked by white and red rectangles in panel {\bf a}, respectively. The data in panels {\bf a}-{\bf d} are measured at $T=\SI{1.5} {K}$. {\bf e}, $\mu$ vs. $n$ at different $E$-field values, in the vicinity of $n=0$. $\Delta_\mathrm{0}$ can be extracted from the 
change in $\mu$ at $n=0$.
}
\label{fig2}
\end{figure}

\begin{figure}[H]
\begin{center}
\includegraphics[width=6.0in]{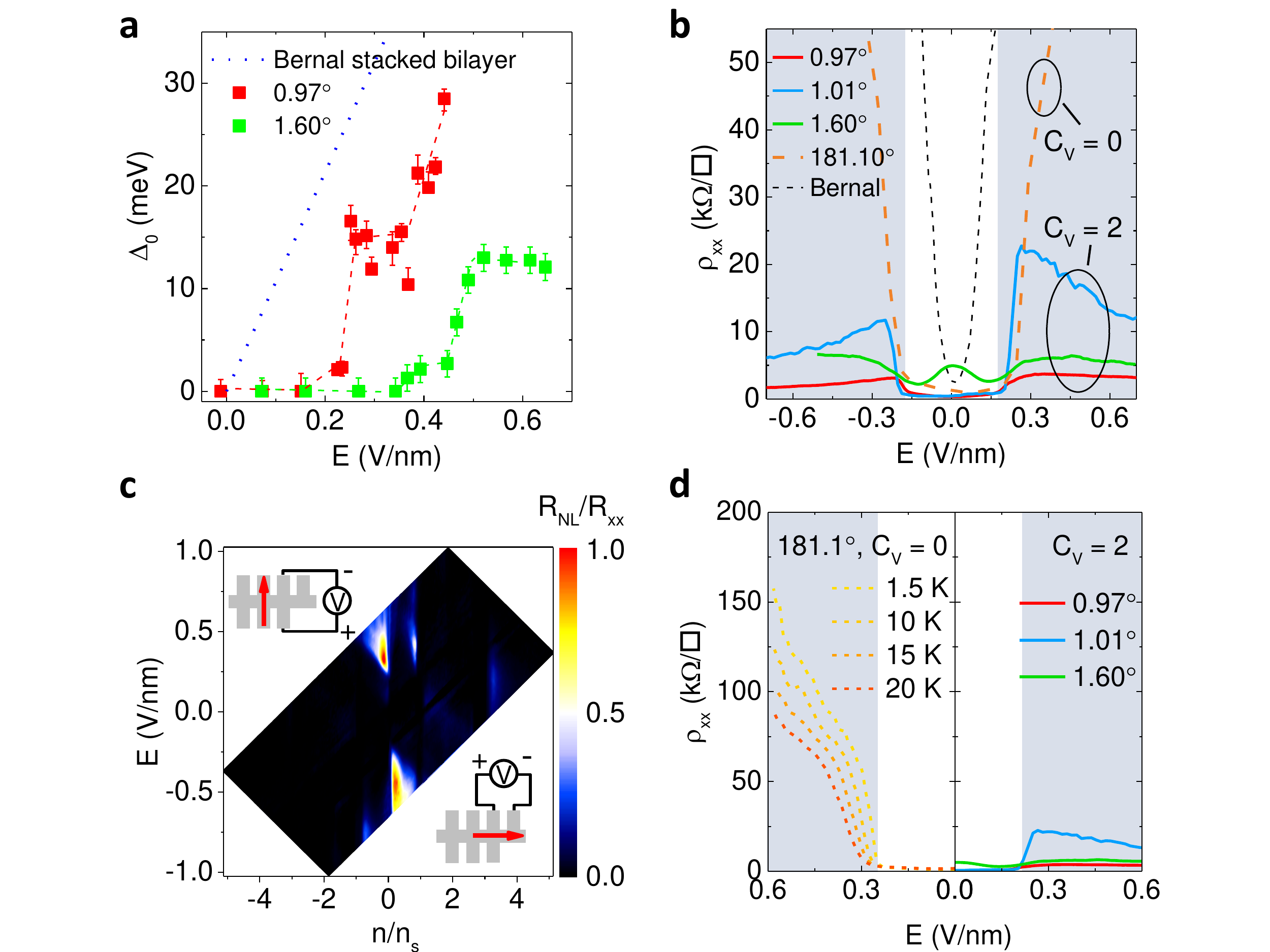}
\end{center}
\caption{{\bf Gapped bulk and edge states. a}, Charge neutrality thermodynamic gap $\Delta_\mathrm{0}$ vs. $E$ for two TDBG samples, and Bernal stacked bilayer graphene. The $\Delta_0$ value in TDBG increases rapidly at a finite $E$, and remains finite once is open. Red and green dashed lines are guides to the eye. The error bars represent measurement uncertainty. {\bf b}, $\rho_\mathrm{xx}$ vs. $E$ at charge neutrality measured in the $\theta=0.97^\circ$, $1.01^\circ$, $1.60^\circ$, and $181.10^\circ$ TDBG samples. The shaded areas mark the gapped region for the $0.97^\circ$ TDBG. In the $0.97^\circ$, $1.01^\circ$, and $1.60^\circ$ samples, the $\rho_\mathrm{xx}$ values show an initial increase when the gap opens, but subsequently decrease with increasing $E$. In contrast, $\rho_\mathrm{xx}$ increases with $E$ to values significantly larger than $h/e^2$ in the $181.10^\circ$ sample. The $\rho_\mathrm{xx}$ vs. $E$ measured at $n=0$ in a Bernal stacked bilayer graphene is included. {\bf c}, Contour plot of $R_\mathrm{NL}/R_\mathrm{xx}$ vs. $n/n_\mathrm{s}$ and $E$ measured in the $\theta=0.97^\circ$ TDBG. The insets show the measurement configuration schematics for $R_\mathrm{NL}$ (upper left) and $R_\mathrm{xx}$ (lower right); the arrows mark the current direction. The large $R_\mathrm{NL}/R_\mathrm{xx} \approx 1$ ratio at $n/n_\mathrm{s}=0$ and finite $E$-fields signals edge transport. {\bf d}, Left panel: temperature dependence of $\rho_\mathrm{xx}$ vs. $E$ at $n=0$ measured in $181.10^\circ$ sample. The shaded area marks the gapped region. Right panel: $\rho_\mathrm{xx}$ vs. $E$ measured in the small-twist-angle TDBGs. The shaded area marks the gapped region for the $0.97^\circ$ sample. The data in panels {\bf a}-{\bf d} are measured at $T=\SI{1.5}{K}$ unless otherwise noted.}
\label{fig3}
\end{figure}

\begin{figure}[H]
\begin{center}
\includegraphics[width=6.3in]{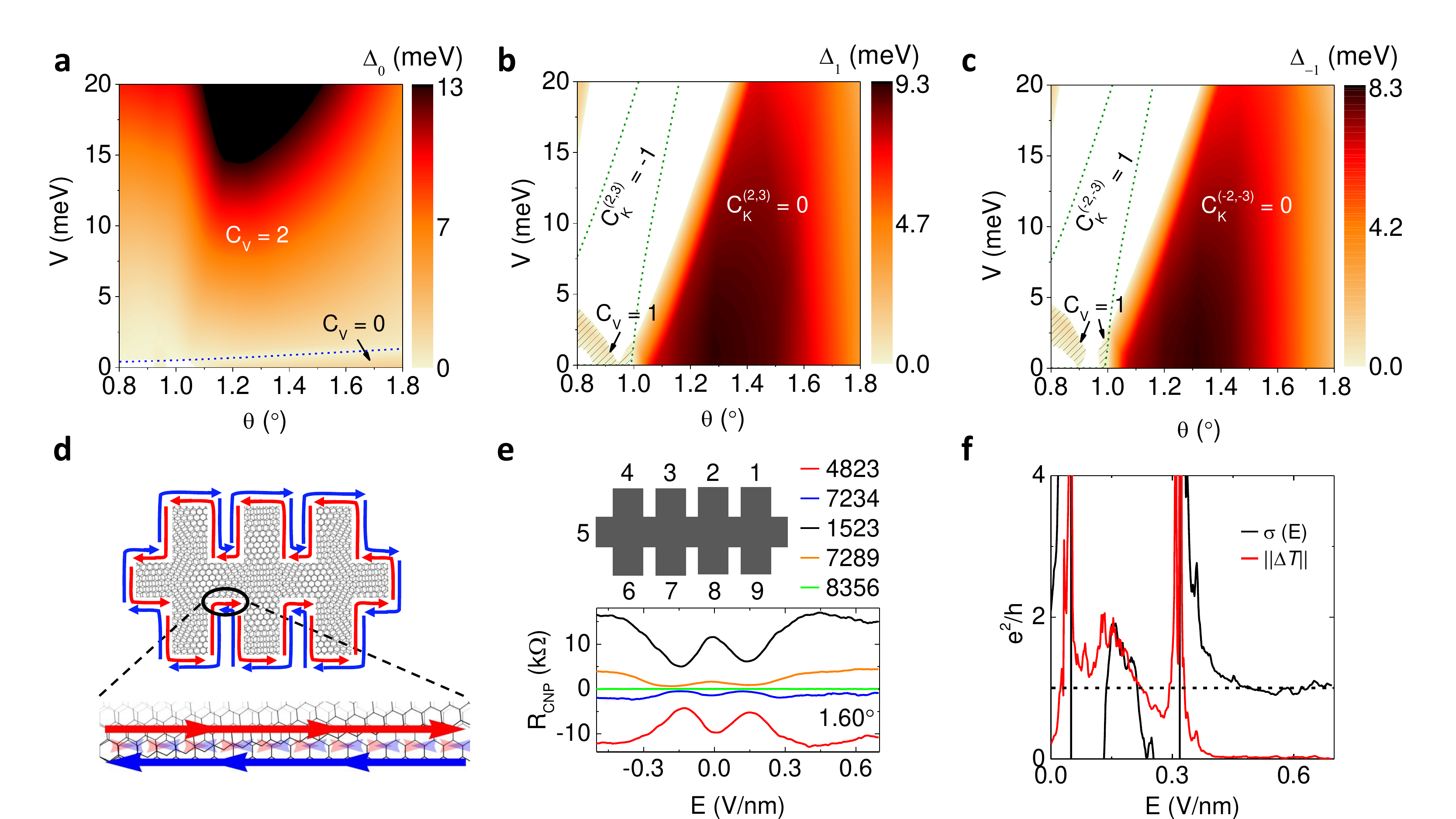}
\end{center}
\caption{{\bf TDBG Phase Diagrams.} {\bf a}-{\bf c}, The valley Chern number $C_V$ as a function of $\theta$ and $V$ for the {\bf a}, gap between the lowest conduction band and highest valence band ($\Delta_0$), {\bf b}, gap between the first and second conduction bands ($\Delta_1$), and {\bf c}, gap between the first and second valence band ($\Delta_{-1}$). The dashed lines indicate Chern number transitions in neighboring bands, e.g. $C_K^{(2,3)}$ ($C_K^{(-2,-3)}$) is the Chern number computed for the connected second and third conduction (valence) bands in the $K$ valley. In panels {\bf b} and {\bf c}, the shaded areas mark gapped regions with nonzero $C_V$ at filling $n/n_{\text{s}} = +1$ and $-1$, respectively. {\bf d}, Schematic of counter-propagating edge states. Due to the breaking of the valley symmetry, scattering on the edge can lead to localization. {\bf e}, Examples of four-terminal resistance at charge neutrality ($R_\mathrm{CNP}$) vs. $E$ measured in a multi-terminal TDBG with $\theta = 1.60^\circ$. The four-digit notations denote the measurement configurations in the form of (current in, current out, voltage $+$, voltage $-$). {\bf f}, $T(E)$ calculated using 64 independent resistance measurements. $\sigma(E)$ diverges when the gap closes, and has values of the order of $e^2/h$ in the topological phase. For $E< \SI{0.33} {V/nm}$, $\sigma$ shows significant variation, expected in a bulk conductor. The norm of $\Delta T(E) = T(E) - T(\SI{0.7} {V/nm})$ is close to zero in the gapped phase, indicating that all $T(E)$ entries are insensitive to the applied $E$-field.}
\label{fig4}
\end{figure}
\clearpage

\section*{Methods}
\textbf{Sample Fabrication} \newline
All the graphene, graphite, and hBN flakes used to fabricate our samples are mechanically exfoliated, and inspected by optical microscopy. The hBN flakes are subsequently inspected with atomic force microscopy to confirm their thickness and surface roughness. Optical contrast and Raman spectroscopy were used to confirm the layer number for monolayer and bilayer graphene. The TDBG samples are assembled by sequential pick-up steps utilizing a hemispherical polypropylene carbonate (PPC)/polydimethylsiloxane (PDMS) handle. The back-gate structure is first prepared by picking up a monolayer or bilayer graphene with hBN, followed by a set of pre-trimmed graphite contacts. The stacked structure is then placed on a SiO$_2$/Si substrate, or a prepared hBN/graphite stack, in the case of samples with graphite substrate gate, to form the bottom structure of the sample. Starting with another large bilayer graphene flake trimmed into two separate sections by lithography and O$_2$ plasma etching, another hBN is used to sequentially picked up the two sections, with a rotation of a small, controlled angle between the two pick-ups, to form the TDBG. The TDBG is then placed on the bottom structure, and a graphite top-gate is place on the TDBG. The structure is then trimmed into a Hall-bar shaped channel using CHF$_3$ and O$_2$ plasma etching, which also creates exposed one-dimensional edges of the TDBG and the graphite contacts to the back-gate. Metal (Cr/Pd/Au) edge contacts are evaporated to finalize the sample.

\noindent\textbf{Measurement Setup} \newline
The samples are measured in a variable-temperature liquid $^4$He cryostat with a base temperature of 1.5 K. Three- and four-point  resistance measurements low-frequency (7 – 13 Hz) lock-in techniques are performed on the TDBG and GrBG layers. Source currents of different frequencies are used on TDBG and GrBG layers, to avoid cross-talk. A radio-frequency transformer is used to flow an AC current in the GrBG layer while applying a DC bias $V_\mathrm{BG}$.

\appendix
\renewcommand{\thefigure}{S\arabic{figure}}
\section*{Supplementary Information}

\section{Magnetotransport Measurements and Hofstadter Butterfly}
\label{app:Hofstadter}

\setcounter{figure}{0}
\begin{figure*}
\begin{center}
\includegraphics[width=5in]{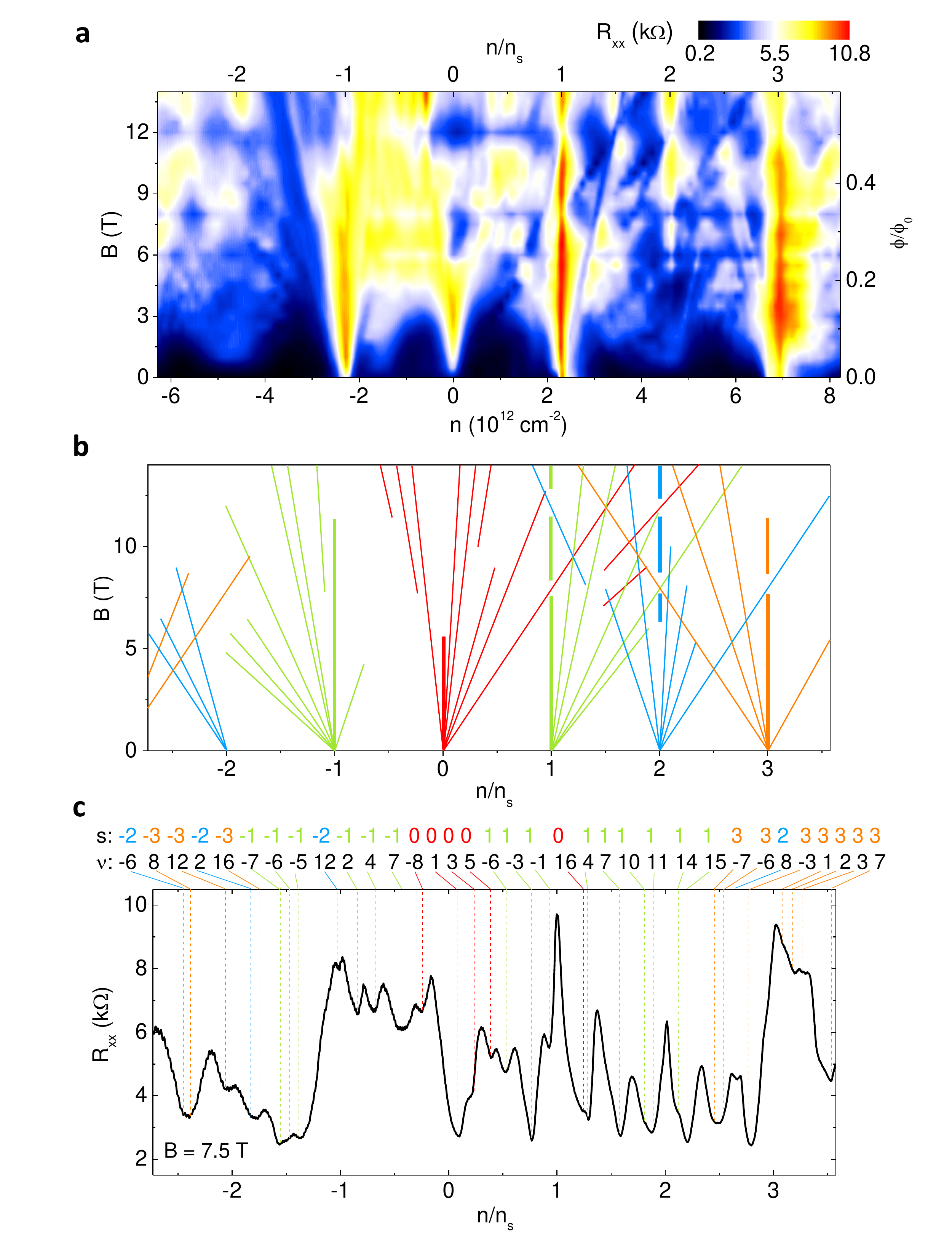}
\end{center}
\caption{{\bf Hofstadter butterfly measured in the $\theta = 0.97^\circ$ TDBG. a}, Contour plot of $R_\mathrm{xx}$ vs. $n$ (bottom axis), $n/n_\mathrm{s}$ (top axis) and $B$; the right axis shows $\phi/\phi_\mathrm{0}$. The data were measured at a fixed $E=\SI{0.1}{\V/\nm}$. {\bf b}, QHSs observed in panel {\bf a}. The fans originating from $n/n_\mathrm{s}=0, \pm1, \pm2, \pm3$ are shown in red, green, blue, and orange, respectively. {\bf c}, $R_\mathrm{xx}$ vs. $n$ measured at $B=\SI{7.5}{T}$, dashed lines denote $s$ and $\nu$ for each resistance minimum; the $s$ indices are labeled with colors corresponding to the fans in panel {\bf b}.}
\label{figS2}
\end{figure*}

\begin{figure*}
\begin{center}
\includegraphics[width=5in]{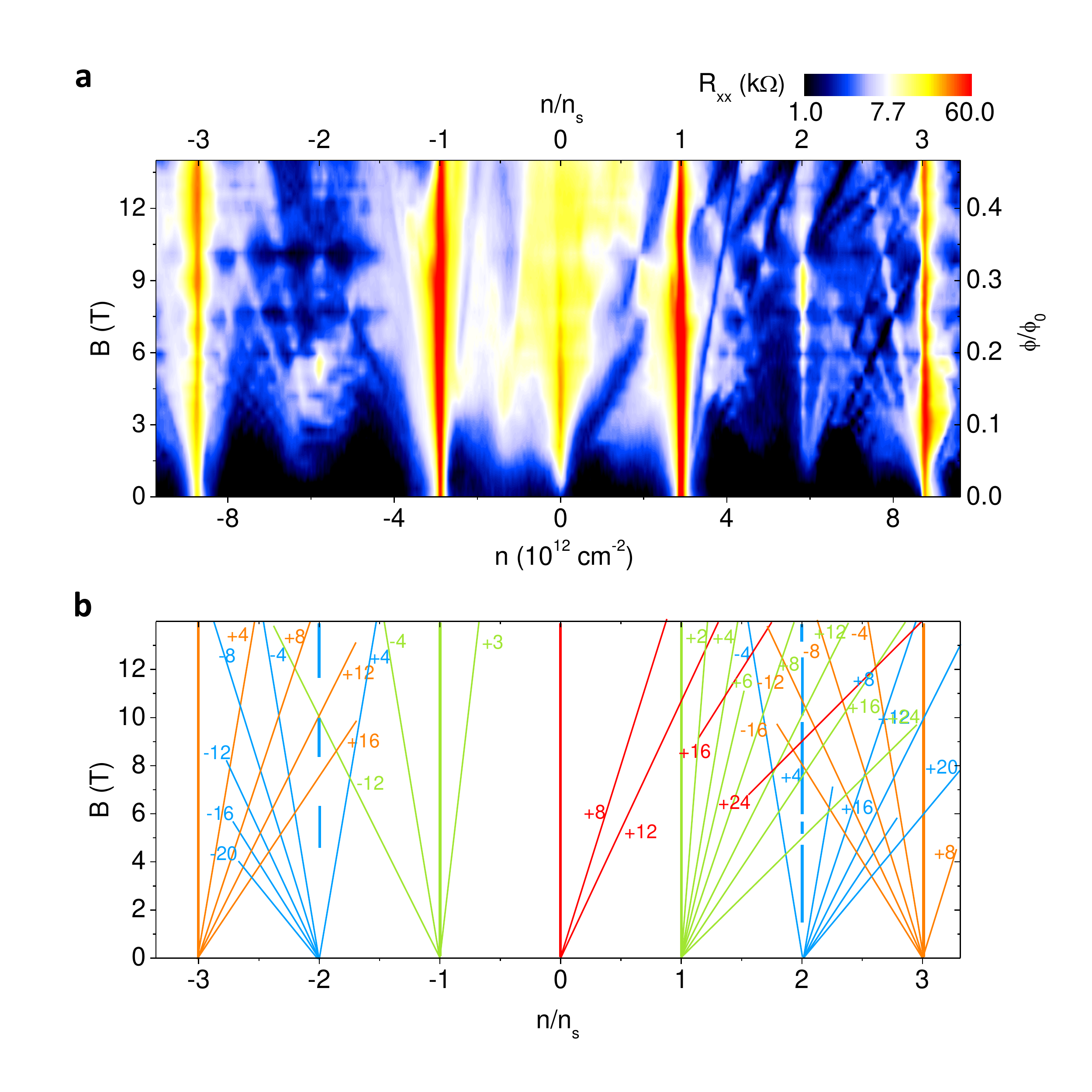}
\end{center}
\caption{{\bf Hofstadter butterfly measured in the $\theta = 181.10^\circ$ TDBG. a}, Contour plot of $R_\mathrm{xx}$ vs. $n$ (bottom axis), $n/n_\mathrm{s}$ (top axis) and $B$; the right axis shows $\phi/\phi_\mathrm{0}$. The data were measured at a fixed $E=\SI{0}{\V/\nm}$. {\bf b}, QHSs observed in panel {\bf a}. The fans originating from $n/n_\mathrm{s}=0, \pm1, \pm2, \pm3$ are shown in red, green, blue, and orange, respectively. The $\nu$ values are labeled for each QHS observed.}
\label{fig180degHof}
\end{figure*}

Magnetotransport measurements are performed for the TDBG samples in perpendicular magnetic field up to $B=\SI{14}{T}$. In Fig. \ref{figS2}a and Fig. \ref{fig180degHof}a, we show $R_\mathrm{xx}$ vs. $n/n_\mathrm{s}$ and $B$, with right axis as $\phi/\phi_\mathrm{0}$, measured in the $\theta=0.97^\circ$ and $181.10^\circ$ samples at constant $E$-fields $E=\SI{0.1}{\V/\nm}$ and $E=\SI{0}{\V/\nm}$, respectively. Here, $\phi$ is the magnetic flux per moir\'e unit cell, $\phi=BA$, $A$ is the moir\'e unit cell area $A=4/n_\mathrm{s}$, and $\phi_\mathrm{0}$ is the magnetic flux quanta $h/e$. The sample is kept at a fixed $E$-field during the measurement by sweeping the top and back-gate biases simultaneously, while keeping $V_\mathrm{TG}/V_\mathrm{BG} = C_\mathrm{BG}/C_\mathrm{TG}$. The ratio $C_\mathrm{BG}/C_\mathrm{TG}$ is determined by the slope of the charge neutrality line from the contour of $R_\mathrm{xx}$ vs. $V_\mathrm{TG}$ and $V_\mathrm{BG}$. For both $\theta=0.97^\circ$ and $181.10^\circ$ samples, a fractal Hofstadter butterfly spectrum is observed. The Hofstadter butterfly consists of trajectories of Landau level gaps in energy spectrum developed under a periodic potential and perpendicular magnetic field, and can be characterized by the Diophantine relation
\begin{equation}
\label{Dpt}
    \frac{n}{n_\mathrm{s}}= \frac{\nu}{4}\frac{\phi}{\phi_\mathrm{0}}+s,
\end{equation}
$\nu$ and $s$ are integers associated with Landau level and moir\'e subband fillings, respectively, and index the QHSs. By fitting the fractal Hofstadter butterfly with Eq. \ref{Dpt}, we are able to determine the carrier densities at each ($V_\mathrm{TG}$, $V_\mathrm{BG}$) point, and confirm the capacitances of the top and bottom gates. In the $\theta=0.97^\circ$ sample, the capacitances are extracted to be $C_\mathrm{TG}=\SI{88.72}{nF/cm^{2}}$ and $C_\mathrm{BG}=\SI{64.76}{nF/cm^{2}}$. We summarize the QHSs observed in Fig. \ref{figS2}a (Fig. \ref{fig180degHof}a) and show them in Fig. \ref{figS2}b (Fig. \ref{fig180degHof}b). The QHS fans originated from $\nu=0$, $\pm 1$, $\pm 2$ and $\pm 3$ are observed. The QSHs with $\nu = 0$, namely at densities with integer $n/n_\mathrm{s}=s$, are marked with vertical lines in Fig. \ref{figS2}b and Fig. \ref{fig180degHof}b. The vertical lines at integer $n/n_\mathrm{s}$ interrupted by QHSs originated from a different $s$-index are marked in Fig. \ref{figS2}b and Fig. \ref{fig180degHof}b, as a signature of nontrivial topology \autocite{PhysRevLett.125.236804}. 

Figure \ref{figS2}c shows $R_\mathrm{xx}$ as a function of $n$ taken at $B=\SI{7.5}{T}$, which shows oscillations associated with well-developed QHSs. We show the $s$ and $\nu$ indices for the resistance minima observed, and note the presence of $\nu$ indices not a multiply of $4$ suggests the lifting of spin and valley symmetry. The observation of plethora of the QHSs testifies to the high quality of our samples.

\section{Chemical Potential Extraction}
\label{app:CPE}

The chemical potentials of TDBG and GrBG can be written in terms of gate biases and capacitances of the heterostructure. We begin by analyzing the band diagram in Fig. \ref{figS1}, where $V_\mathrm{\epsilon,T}$, $V_\mathrm{\epsilon,B}$, and $V_\mathrm{\epsilon,s}$ are the changes in electrostatic potential across the top, back, and substrate dielectrics, respectively. $n_\mathrm{s}$ is the carrier densities in the substrate gate, and $\mu_\mathrm{BG}$ is the chemical potential of the GrBG. An applied $V_\mathrm{TG}$ is the sum of the potential drop across the top-gate dielectric and the chemical potential of the TDBG:

\begin{equation}
eV_{\mathrm{TG}}=eV_{\mathrm{\epsilon,T}}+\mu,
\end{equation}
where $e$ is the electron charge. Similarly, $V_\mathrm{BG}$ and $V_\mathrm{s}$ can also be written as:
\begin{equation}
eV_{\mathrm{BG}}=eV_{\mathrm{\epsilon,B}}+\mu-\mu_\mathrm{BG},
\end{equation}
\begin{equation}
eV_{\mathrm{s}}=eV_{\mathrm{BG}}+eV_{\mathrm{\epsilon,s}}+\mu_\mathrm{BG},
\end{equation}
while the potential drops across the dielectrics can be written as
\begin{equation}
C_{\mathrm{TG}}V_{\mathrm{\epsilon,T}}=e(n+n_\mathrm{BG}+n_\mathrm{s}),
\end{equation}
\begin{equation}
C_{\mathrm{BG}}V_{\mathrm{\epsilon,B}}=-e(n_\mathrm{BG}+n_\mathrm{s}),
\end{equation}
\begin{equation}
C_{\mathrm{s}}V_{\mathrm{\epsilon,s}}=-en_\mathrm{s},
\end{equation}
here, $C_{\mathrm{TG}}$, $C_{\mathrm{BG}}$, and $C_{\mathrm{s}}$ are the gate capacitances of the top, back and substrate gates, respectively. Note that during all the measurements, the potential of the TDBG is always kept at ground. Combining the six equations above, the following relations can be obtained:

\begin{figure}[t]
\begin{center}
\includegraphics[width=4.6in]{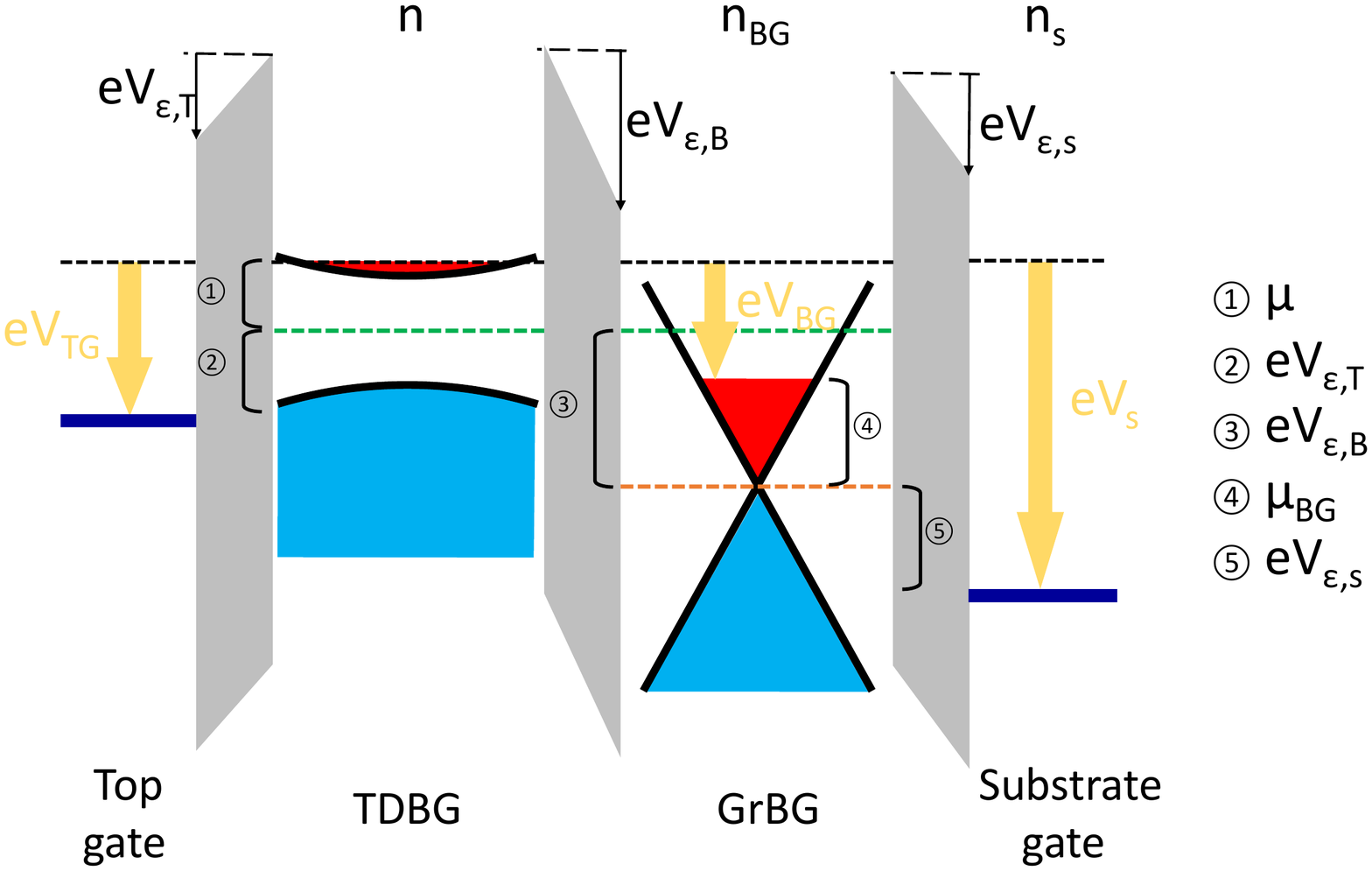}
\end{center}
\caption{Band diagram of the TDBG heterostructure. The TDBG is kept at the ground potential (black dashed line). The green and orange dashed lines mark the charge neutrality of TDBG and GrBG, respectively.}
\label{figS1}
\end{figure}

\begin{equation}
C_{\mathrm{TG}}V_{\mathrm{TG}}=e(n+n_\mathrm{BG})+C_\mathrm{s}(V_{\mathrm{BG}}-V_{\mathrm{s}})+\frac{\mu_\mathrm{BG}C_\mathrm{s}}{e}+\frac{\mu C_\mathrm{TG}}{e},
\end{equation}
\begin{equation}
C_{\mathrm{BG}}V_{\mathrm{BG}}=-en_\mathrm{BG}-C_\mathrm{s}(V_{\mathrm{BG}}-V_{\mathrm{s}})-\frac{\mu_\mathrm{BG}C_\mathrm{s}}{e}+\frac{(\mu-\mu_\mathrm{BG})C_\mathrm{BG}}{e}.
\end{equation}
At GrBG charge neutrality, $n_\mathrm{BG}=0$ and $\mu_\mathrm{BG}=0$ are satisfied, the relations above become:
\begin{equation}
C_{\mathrm{TG}}V_{\mathrm{TG}}=en+C_\mathrm{s}(V_{\mathrm{BG}}-V_{\mathrm{s}})+\frac{\mu C_\mathrm{TG}}{e},
\end{equation}
\begin{equation}
C_{\mathrm{BG}}V_{\mathrm{BG}}=-C_\mathrm{s}(V_{\mathrm{BG}}-V_{\mathrm{s}})+\frac{\mu C_\mathrm{BG}}{e},
\end{equation}
we then obtain the chemical potential of TDBG in terms of ($V_{\mathrm{BG}}$, $V_{\mathrm{s}}$, $C_{\mathrm{BG}}$, $C_{\mathrm{s}}$) at GrBG charge neutrality,
\begin{equation}
\label{chemical potential}
    \mu=eV_\mathrm{BG} \left( 1+\frac{C_\mathrm{s}}{C_\mathrm{BG}}\right)-eV_\mathrm{s} \frac{C_\mathrm{s}}{C_\mathrm{BG}}.
\end{equation}
We note that, according to Eq. \ref{chemical potential}, the resolution of the chemical potential measurement is limited by the accuracy in determining the back-gate charge neutrality as a function of $V_\mathrm{BG}$, $\delta \mu=(1+C_\mathrm{s}/C_\mathrm{BG})\delta V_\mathrm{BG}$.

\section{Additional Data}
\label{app:CNP res}

\begin{figure}[H]
\begin{center}
\includegraphics[width=3.5in]{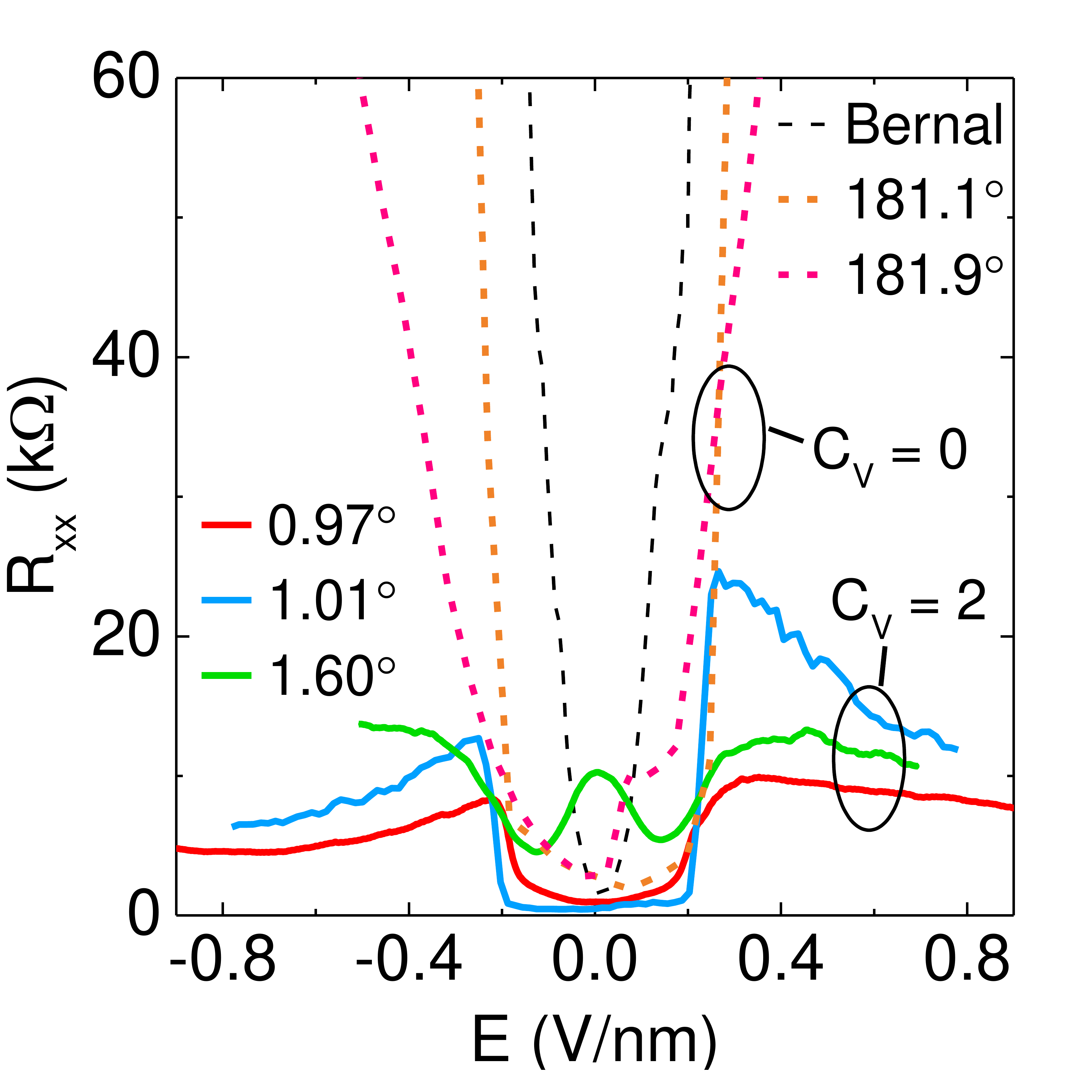}
\end{center}
\caption{$R_\mathrm{xx}$ vs. $E$ at charge neutrality measured in TDBG samples with $\theta=0.97^\circ$, $1.01^\circ$, $1.60^\circ$, $181.1^\circ$ and $181.9^\circ$, and Bernal stacked bilayer graphene \autocite{lee2014}.}
\label{CNP_res}
\end{figure}

Figure \ref{CNP_res} shows charge neutrality resistance ($R_\mathrm{xx}$) vs. $E$-field data measured in TDBGs with $\theta=0.97^\circ$, $1.01^\circ$, $1.60^\circ$, $181.1^\circ$ and $181.9^\circ$. The data shows that the charge neutrality resistance in the TDBGs with $\theta \approx 180^\circ$ increases with the $E$-field and reaches values larger than $h/e^2$, supporting the findings that the TDBG with a twist angle $\theta \approx 180^\circ$ is a band insulator at charge neutrality at finite $E$-field, in contrast with the TDBG with $\theta \approx 0^\circ$. For comparison, $R_\mathrm{xx}$ vs. $E$ data for a Bernal stacked bilayer graphene is included.

\section{Theoretical Calculation of the Topological Phase Diagram}
\label{app:PD}

In this Appendix, we briefly describe the band structure calculations and determination of the phase diagrams, giving the Chern number and gap as a function of $\theta$ and $E$. Our starting point is the Bistritzer-MacDonald Hamiltonian considered in Ref. \autocite{burg2020}. To summarize, the TDBG band structure is computed across a moir\'e Brillouin Zone on the scale of $k_M =\frac{8\pi}{3} a^{-1} \sin \frac{\theta}{2} $ where $\theta$ is the moir\'e twist angle and $a = \SI{2.46}{\angstrom}$ is the graphene lattice constant. The Bistritzer-MacDonald Hamiltonian, per spin per valley, is written as a honeycomb  continuum model in momentum space  by plane-wave expansions which we truncate to five shells of moir\'e reciprocal lattice vectors. We do not employ the particle-hole symmetric approximation in order to uncover finer details of the phase diagram. 

The band gaps shown in Figs. 4a-c are determined by computing the band structure on a fine mesh of angle $\theta$ (from $0.8^\circ$ to $1.8^\circ$), and inter-layer potential $V$ (from $0$ to $\SI{20} {meV}$). We note that the conversion between $E$ and $V$, $V=s e E d$ depends on a phenomenological screening parameter $s$. Here, $e$ is the electron charge and $d= \SI{3.4}{\text{\AA}}$ is the inter-layer distance. A quantitative understanding of screening in TDBG is beyond the scope of the study.

As shown in Fig. 1f, the low energy bands are labeled in a convention where valence bands at charge neutrality are indexed with negative numbers, e.g. band $(-2)$ is the second band below charge neutrality, and the conduction bands are indexed with a positive number, e.g. band $(1)$ is the upper flat band just above charge neutrality. To compute the Chern number over a single band (or set of bands), we use the formula \autocite{2007PhRvL..98d6402B}
\bea
\label{eq:chern}
C^{(n)} = \frac{3 \sqrt{3}}{2} \times \frac{i}{2\pi \epsilon^2 N} \sum_{k_x,k_y} \text{Tr } \mathcal{P}(k_x, k_y) [\mathcal{P}(k_x + \epsilon, k_y) - \mathcal{P}(k_x, k_y), \mathcal{P}(k_x, k_y+\epsilon) - \mathcal{P}(k_x, k_y)]
\eea
where $\mathcal{P}(k_x, k_y) = U(k_x, k_y)U^\dag(k_x, k_y)$ is the projection matrix on the band $n$, i.e. $U(k_x, k_y)$ is the eigenvector of band $n$, $(k_x,k_y)$ are the coordinates of the moir\'e Brillouin zone which has area $\frac{3 \sqrt{3}}{2} k_M^2$, $\epsilon = 10^{-4}$ approximates the numerical derivative, and $N=16899$ is the number of points in hexagonal moir\'e BZ making up the momentum sum. Eq.~(\ref{eq:chern}) is written in terms of projectors and hence is manifestly gauge-invariant. We take $N$ to be large so that very good quantization is observed in the calculated Chern numbers. To calculate the Chern number of a connected set of bands, $U(k_x, k_y)$ should be understood as a matrix whose columns are the eigenvectors of the bands under consideration, thereby generalizing the one band example. By computing the Chern numbers of the occupied bands at a giving filling, we determine the total Chern number in the $K$ valley by
\bea
C_K = \sum_{n \text{ occupied}} C^{(n)}
\eea
and, by time-reversal symmetry $C_{K'} = - C_K$, so finally
\bea
C_V = \frac{C_K - C_{K'}}{2} \ . 
\eea

We now discuss the phase diagrams obtained in this manner in Figs. 4a-c. At charge neutrality (Fig. 4a) we find that for small $V$ there is a gapped region where the system is a trivial insulator with $C_V = 0$. While the experimental results do not show a gap at $E=0$, its theoretical value is below the resolution limit of our thermodynamic chemical potential measurements. 

A line of gap closings (dashed in blue in Fig. 4a) separates this region from the nontrivial phase of interest in this work where $C_V = 2$. We note that the bands which carry the nontrivial Chern numbers change within this region, which we show with green dashed lines in Figs. 4b-c. For larger angles, the highest valence band $(-1)$ carries $C^{(-1)} =2$, whereas for smaller angles, a band-touching transition \emph{within the occupied manifold} separates a region where $C^{(-1)} =1$ and $C^{(-2,-3)} =1$, so the Chern number is split over the highest three valence bands. At charge neutrality, there is no difference between these phases, but when both ($n/n_{\text{s}} = +1$) or neither ($n/n_{\text{s}} = -1$) of the flat bands are filled, this transition changes the observable valley Chern number. Let us first discuss Fig. 4b where the upper flat band is occupied ($n/n_{\text{s}} = +1$). The shaded region in this plot highlights a gapped phase where the valley Chern number $C_V=2$ of the bands below charge neutrality is partially canceled by the Chern number $C_K^{(1)} = -1$ of the upper flat band, leading to $C_V = 2-1 = 1$. At larger angles where $C_K^{(1)} = -2$, there is a complete cancellation and $C_V = 0$. The analogous situation appears for filling $n/n_{\text{s}} = -1$ shown in Fig. 4c where for small angles $C_K^{(-2,-3)} = 1$ and hence $C_V = 1$. For larger angles, $C_K^{(-2,-3)} = 0$ and the occupied band is trivial. However, the shaded region in this case is interrupted by a gap closing due to the small particle-hole symmetry breaking. 

\begin{figure}[H]
\begin{center}
\includegraphics[width=6in]{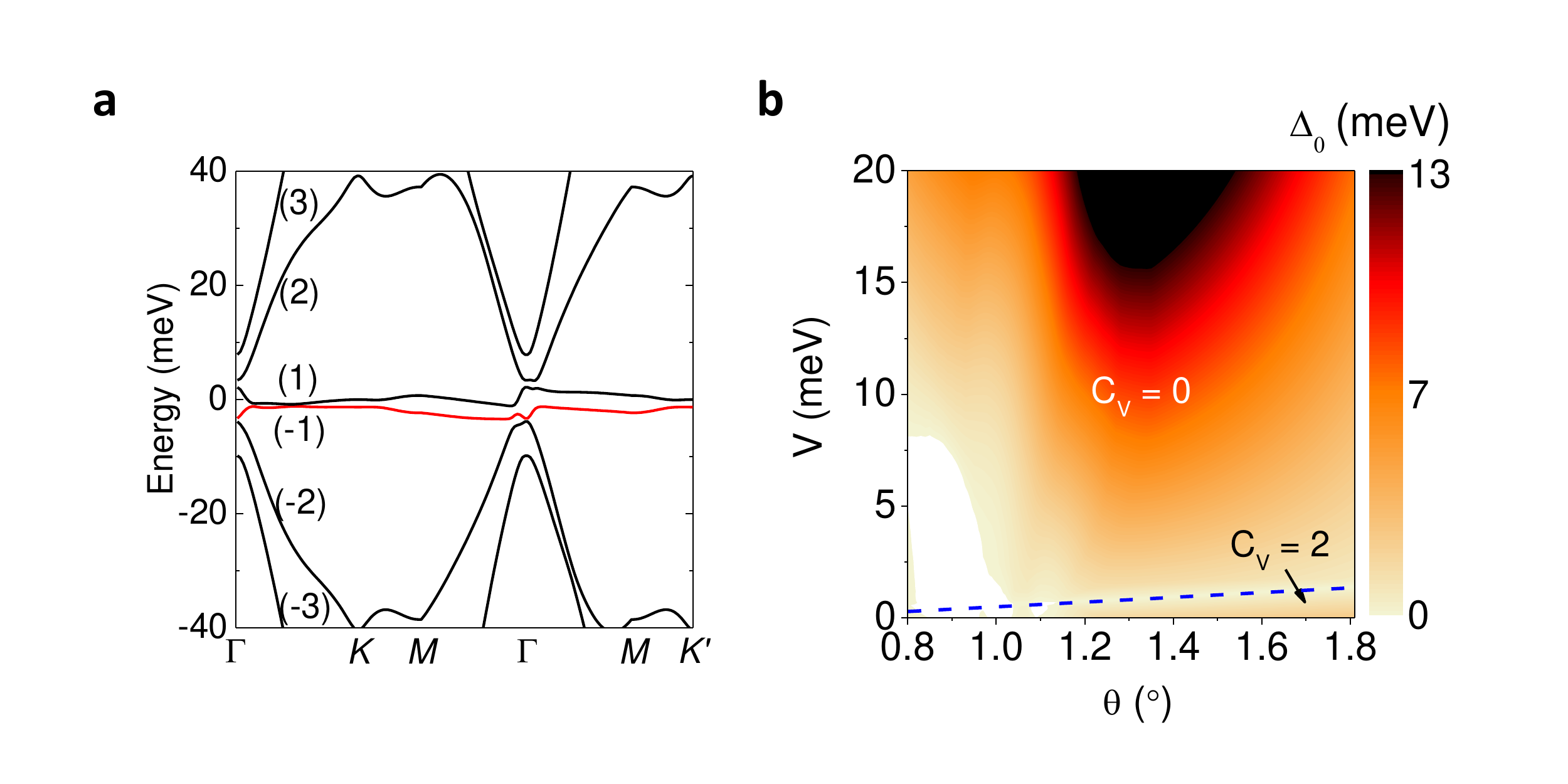}
\end{center}
\caption{{\bf a}, Band structure of TDBG at $181.10^\circ$.We highlight the pair of flat bands near the Fermi level and the pair of passive bands with large dispersion, as is also seen in TDBG near $1^\circ$. The bands are numbered as in Fig. \ref{fig1}f. {\bf b}, The valley Chern number $C_V$ and indirect gap $\Delta_0$ as a function of twist angle $\theta$ and layer potential difference $V$. $\Delta_0$ is the gap between the lowest conduction band and highest valence band, the blue dashed line marks where the direct gap at charge neutrality closes. }
\label{180bs}
\end{figure}

We now discuss the band structure and topology of the $181.10^\circ$ TDBG device, which is AB-BA stacked due to the additional $180^\circ$ rotation. The details of the system can be found in Ref. \autocite{chebrolu2019} and Ref. \autocite{koshino2019}, as well as a discussion of the topology. The salient features for this work are (1) the moir\'e band structure is very similar to small-angle TDBG, featuring two nearly flat bands at charge neutrality and a pair of energetically connected passive bands below and above shown in Fig. \ref{180bs}a, and (2) the topology of the $180^\circ$ system is very different, providing a strong point of comparison. We show the full phase diagram in Fig. \ref{180bs}b calculated using Eq. \ref{eq:chern}. The most important property of the system is its behavior in large applied $E$-field at charge neutrality where the system is a \emph{trivial} insulator with $C_V = 0$. This is in stark contrast to the phase diagram of $1.60^\circ$ TDBG, shown in Fig. 4a, where the system is a nontrivial valley Chern insulator in large $E$ field. An experimental comparison of these systems shows different transport behavior, which we associate with the differences in band topology.

\section{Landauer-B\"{u}ttiker Formalism}
\label{app:LB}

In this Appendix, we review the Landauer-B\"{u}ttiker formalism and highlight how the charge conservation and gauge invariance are taken into account. We then present our exact inverse which relates measured resistances to the transmission matrix that quantitatively describes the edge states for an arbitrary number of terminals. We also define the conductance order parameter $\sigma$ and discuss its properties. This analysis is general and can be directly applied to nonlocal transport measurements in any system. Finally, we describe the analysis of the nonlocal measurements in our TDBG samples and give quantitative evidence for topologically protected edge states. 

\subsection{Introduction}

\begin{figure*}
 \centering
 \includegraphics[width=5cm]{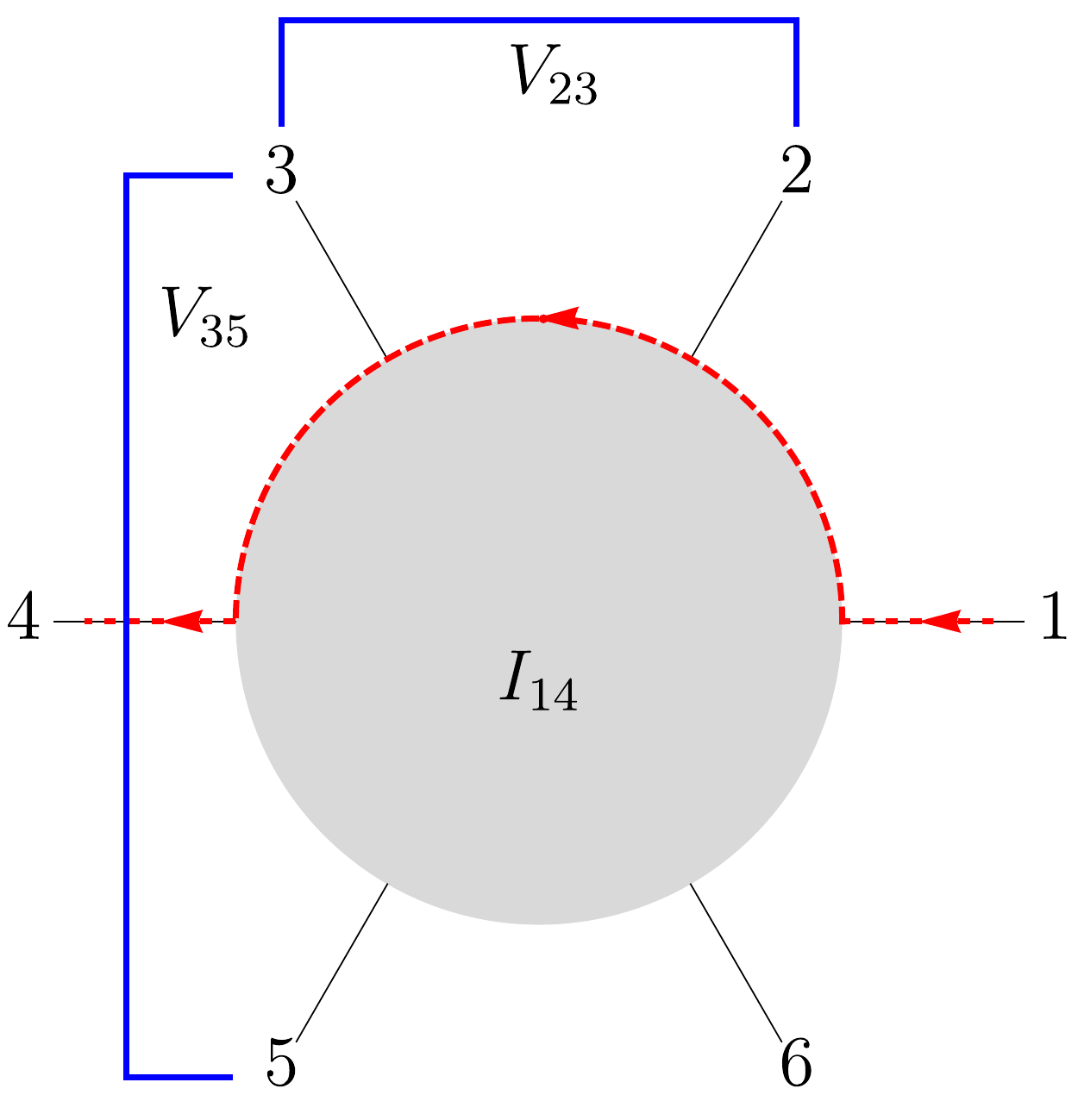}
\caption{An example of a device with $N=6$. We show the current in red with $I_1 = - I_4$ and all other components of $I_i=0$. We depict two voltage measurements $V_{35} = V_5 - V_3$ and $V_{23} = V_3 - V_2$. } 
\label{deviceLB}
\end{figure*}

We consider a system of $N$ electrodes which may be connected to conventional electronic voltage $V_i$ or current $I_i$ leads, as depicted in Fig. \ref{deviceLB}. In the Landauer-B\"{u}ttiker (LB) formalism\autocite{5390001,PhysRevLett.57.1761,PhysRevB.38.9375}, the measured current and voltage are related by the transmission probabilities $T_{ij}$, where $T_{ij}$ is the probability of transport from $i \leftarrow j$, in the form
\bea
\label{LBeq}
I_i &= \frac{\mathcal{N} e^2}{h} \sum_{j=1}^N (T_{ji} V_i - T_{ij} V_j)\\
\eea
where $\mathcal{N}$ is the number of edge states of a given chirality. For the case of a valley Chern insulator with spin degeneracy, we have $\mathcal{N} = 2C_V$. In experiments, the current is injected into one terminal and exits through another, so the vector $I_i$ is nonzero and opposite for two entries. Using a voltimeter, differences in the entries of $V_i$ can be measured between any pair of terminals. 

The LB equation Eq. \ref{LBeq} is essentially Ohm's law on the edge, saying that the current is proportional to the voltage drops weighted by the transmission probabilities. As an example, consider a perfectly clean quantum hall system with $C=1$, yielding a single chiral edge state. Because the state is chiral, electrons are only transported in one direction around the sample, and they are dissipationless. Hence $T_{ij} = \delta_{i,j+1}$ where $i,j$ are taken mod $n$ to be periodic around the sample. In this case, Eq. \ref{LBeq} gives
\bea
I_i &= \frac{e^2}{h} \sum_{j=1}^N (\delta_{j+1,i} V_i - \delta_{i,j+1} V_j) = \frac{e^2}{h} (V_i - V_{i-1}) \\
\eea
From Ohm's law, $\Delta V = I R$, we find a resistance of $R = h/e^2 = \SI{25.81} {k\Omega}$.

\subsection{Gauge Invariance and Current Conservation}
\label{app:gicc}

The LB equation is subject to current conservation and gauge invariance. Current conservation means that the physical currents $I$ must satisfy $\sum_{j=1}^N I_j = 0$, and gauge invariance means that the LB equation must be invariant under $V_j \to V_j + \text{const}$. Current conservation means that physical currents $I$ occupy an $n-1$ dimensional subspace of $\mathbb{R}^n$ subject to $\sum_{j=1}^N I_j = 0$, and gauge invariance gives an equivalence relation $V_i \to V_i + \text{const}$. that reduces physically distinct voltage configurations to an $n-1$ dimensional subspace of $\mathbb{R}^N$.
Consider Eq. \ref{LBeq} for $V_i = v= \text{const}$. which corresponds to every terminal being at the same voltage, so no current flows and $I_i = 0$. Eq. \ref{LBeq} then reads
\bea
\label{eq:Tijcon}
0 &= \sum_{j=1}^N (T_{ji} - T_{ij} ) v \\
\eea
so the $i$th row sum is equal to the $i$th column sum. Eq. \ref{eq:Tijcon} ensures that $V_i$ and $V_i + v$ are physically indistinguishable, which is a requirement of gauge invariance. We now show that for the LB equation to be consistent, there is gauge invariance if there is current conservation. We can rewrite the LB equation in matrix notation as
\bea
\label{eq:Smat}
I = \frac{2 C_V e^2}{h} S V
\eea
where $S_{ij} = \delta_{ij} \sum_{k=1}^N T_{ik} - T_{ij}$. Note that the sum of the elements of each row of $S$ is equal to zero. Hence, $S$ has a nontrivial nullspace which is one-dimensional and is spanned by the vector $\mu^T = (1,1,\dots, 1)$. This is due to gauge invariance $SV = S(V+\mu)$, which reads
\bea
S \mu = 0 \ .
\eea
Note that current conservation may be written as $\mu^T I = 0$, which is consistent with the LB equation because
\bea
\mu^T I \propto \mu^T S V= 0 \ . 
\eea
Here we have used that
\bea
\mu^T S &= \sum_{i=1}^N \mu_i S_{ij} =  \sum_{i=1}^N (T_{ji} - T_{ij}) = 0 
\eea
so $\mu$ is in the left and right nullspaces. Note that time reversal symmetry (TRS) is sufficient to show that the left and right nullspaces are equal because TRS implies $T_{ij} = T_{ji}$, and hence $S = S^T$. We will assume TRS holds in our application to TDBG at charge neutrality. 

Finally, we discuss the diagonal entries of $T_{ij}$, which correspond to on-site transmissions. We will show that these are not physically meaningful quantities, as is perhaps intuitive. To do so, we study the LB equation under the transformation $T_{ij} \to T_{ij} + t_i \delta_{ij}$ 
\bea
I_i  = \frac{2 C_V e^2}{h} \sum_{j=1}^N (T_{ji} V_i - T_{ij} V_j) &\to \frac{2 C_V e^2}{h} \sum_{j=1}^N (T_{ji} V_i - T_{ij} V_j) + \frac{2 C_V e^2}{h} \sum_{j=1}^N (t_j \delta_{ji} V_i - \delta_{ij} t_i V_j) \\
&= \frac{2 C_V e^2}{h} \sum_{j=1}^N (T_{ji} V_i - T_{ij} V_j) + \frac{2 C_V e^2}{h} (t_i V_i - t_i V_i) \\
&= I_i \\
\eea
finding that the physical object -- the current -- is invariant. Hence the diagonal terms of $T_{ij}$ are also unobservable. 

To summarize, we have found imposing the physical constraints of gauge invariance and current conservation are equivalent to linear algebraic conditions on the $S$ matrix nullspace which will be crucial to performing the inverse, as we now discuss. 

\subsection{Four Terminal Resistances}

In experiments, a current flows between two terminals, corresponding to $I_i$ being nonzero and opposite in two of its entries and zero otherwise. By placing voltage probes on other pairs of terminals, we can measure four-terminal resistances. It is possible to use the $T$ matrix to calculate voltages in other current configurations as well. Let us describe these measurements in terms of our formalism, assuming $T$ is known. In the next section, we will solve the inverse problem: how one determines $T$ through resistance measurements. 

If $T$ and $I$ are known, we simply need to compute $V$ to determine the resistances. However, because $S\mu = 0$, $S^{-1}$ does not exist. Instead, we can use the pseudo-inverse, or Moore-Penrose inverse\autocite{10.2307/2949777}, $S^+$ to get solutions for $V$ up to a gauge choice. As a brief reminder, $S^+$ can be straightforwardly computed from the singular value decomposition $S = U D V$ where $U,V$ are $N\times N$ orthogonal matrices and $D$ is a diagonal matrix with $\text{rank}(S) = N-1$ nonzero positive entries known as the singular values. Define the diagonal matrix $D^+$ by $D^+_{ii} = 1/D_{ii}$ if $D_{ii} \neq 0$ and $D^+_{ii} = 0$ otherwise. Then $S^+ = V^T D^+ U^T$.

Recall that $S^+ S$ is a projector onto the orthogonal complement of the nullspace of $S$, i.e. it projects out $\mu$. Hence, if we choose a gauge where $\mu^T V = 0$, i.e. $V$ is not in the nullspace, then
\bea
S^+ I &= \frac{2 C_V e^2}{h} S^+ S V = \frac{2 C_V e^2}{h} V \ . \\
\eea
Note that this gauge choice is always possible. For generic $V$ we can make the gauge transformation $V \to V - \frac{\mu^T V}{N} \mu$ which satisfies $\mu^T V = 0$ by construction. 

The four terminal measurement $R_{ij,kl}$ corresponds to driving a current $I$ into terminal $i$ and out of terminal $j$, and measuring the voltage $V_k - V_l$. We write this as $I_m = I e^{ij}_m$ where $e^{ij}$ is a vector whose $m$th component is $e^{ij}_m = (\delta_{im} - \delta_{jm})$ and $I$ is the magnitude of the current. (Note that raised and lowered indices are equivalent here.) As an example in the six terminal system of Fig. \ref{deviceLB}, $e^{14} = (1,0,0,-1,0,0)$. 

Using the pseudo-inverse, we find
\bea
\label{eq:Rijkl}
\frac{2 C_V e^2}{h} (V_{k} - V_l) = I (S^+ e^{ij})_k - I (S^+ e^{ij})_l \\
R_{ij,kl} = \Big( (S^+ e^{ij})_k - (S^+ e^{ij})_l \Big) \frac{h}{2 C_V e^2} \ .
\eea
We remark that $T$ and $S$ are unitless, and $R_{ij,kl}$ as written is proportional to the von Klitzing constant $h/e^2$. 

Note that $R_{ij,kl}$ is gauge-invariant under $V_i \to V_i + \text{const}$, as it must be because it is an observable. From this equation, the anti-symmetries imply $R_{ij,kl} = - R_{ji,kl} = -R_{ij,lk}$. Because $R_{ij,kl}$ is linear, it has simple composition properties:
\bea
R_{ij',kl} + R_{j'j,kl} &= R_{ij,kl} \\
R_{ij,kl'} + R_{ij,l'l} &= R_{ij,kl} \ . \\
\eea
These correspond to Kirchoff's rules. This means there are only $(n-1)^2$ independent components of $R_{ij,kl}$. This is because the gauge invariant part of $S$ is an $(N-1) \times (N-1)$ matrix corresponding to the column and row space, e.g. neglecting the one-dimensional nullspace. We can also do this counting for $T_{ij}$. There are $N^2$ components, but the $N$ diagonal components are unobservable, and there are $N-1$ row/column sum constraints. Indeed, $N^2 - N - (N-1) = (N-1)^2$. 

We return to the six terminal quantum Hall example briefly to illustrate this formalism. The conductance matrix is given by
\bea
S &= \left(
\begin{array}{cccccc}
 1 & 0 & 0 & 0 & 0 & -1 \\
 -1 & 1 & 0 & 0 & 0 & 0 \\
 0 & -1 & 1 & 0 & 0 & 0 \\
 0 & 0 & -1 & 1 & 0 & 0 \\
 0 & 0 & 0 & -1 & 1 & 0 \\
 0 & 0 & 0 & 0 & -1 & 1 \\
\end{array}
\right) \ .
\eea
$S^+$ can be computed with a standard software package such as MATLAB or Mathematica. In fact, $S$ is diagonalizable, so the singular value decomposition is merely an eigenvalue decomposition and can be computed analytically. We obtain
\bea
S^+ &= \frac{1}{12} \left(
\begin{array}{cccccc}
 5 & -5 & -3 & -1 & 1 & 3 \\
 3 & 5 & -5 & -3 & -1 & 1 \\
 1 & 3 & 5 & -5 & -3 & -1 \\
 -1 & 1 & 3 & 5 & -5 & -3 \\
 -3 & -1 & 1 & 3 & 5 & -5 \\
 -5 & -3 & -1 & 1 & 3 & 5 \\
\end{array}
\right)
\eea
so recalling that $S^+ I = e^2/h V$, we can compute the voltages. For concreteness, consider the case in Fig. \ref{deviceLB} where current is injected across the sample, i.e. $I = (1,0,0,-1,0,0) = e^{14}$. Then $V = \frac{h}{e^2}(1,1,1,-1,-1,-1)/2$, and the 4 terminal resistance $R_{14,kl}$ is nonzero, taking value $h/e^2$ only if $k \in \{1,2,3\}$ and $l \in \{4,5,6\}$ which corresponds to taking measurements between the current leads. Any measurement between the current leads will yield the same quantized value $h/e^2$. 

More generally, it is trivial for a computer to calculate $R_{ij,kl}$ for any given $T_{ij}$ using Eq. \ref{eq:Rijkl}. However, in experiments we are faced with the opposite problem. We need to determine an unknown $T$ from a number of $R_{ij,kl}$ measurements. B\"uttiker solved this problem in the case of four terminals, and we will generalize his result to a device of any number of terminals. Notably, this allows us to place very stringent constraints on the edge physics without making any assumptions on the form of $T_{ij}$, as is often done. 

\subsection{Algorithm for the Determination of $T_{ij}$}

We now want to design an efficient and exact inverse algorithm that returns the $T$ matrix from the measured four-terminal resistances. To start, we observe that in terms of the four-terminal resistance, Eq. \ref{eq:Rijkl} may be written as
\bea
R_{ij,kl} = e_{kl}^T S^+ e_{ij} \times \frac{2 C_V e^2}{h}\ . \\
\eea
Physically, $S^+$ is an undetermined matrix corresponding to the inverse conductance, and our measurements $R_{ij,kl}$ correspond to certain overlaps of this matrix given by vectors $e_{ij}$ and $e_{kl}$.
We now describe a procedure of choosing these vectors that allows us to determine $S^+$. In practice, we absorb the coefficient $2C_V$ into the $T$ matrix so that it may be determined empirically from the column sums. Hence our method makes no assumptions on the form of the $T$ matrix and is entirely empirical so no optimization or regression is required.

We now choose a basis of $\mathbb{R}^N$ for the currents and voltages. A simple basis is, for instance, $\{e^{12}, e^{23},\dots, e^{N-1,N}, \mu\}$ where the first $N-1$ vectors correspond to physical measurements, and the last vector $\mu$ spans the nullspace and by definition $S^+\mu = 0$. In experiments, it is helpful to choose a different basis with a minimal number of three terminal measurements, i.e. where the current probes and voltage probes touch the same terminal. This is because three terminal measurements also pick up contact resistance, which can be estimated but does introduce some error. In practice, we find that the contact resistances can be estimated to acceptable tolerances. 

In our nine-terminal device at $\theta = 1.60^\circ$, we choose a current basis that corresponds to injecting current across the sample (in a bulk conductor, this would lead to the deepest penetration of current into the material) and a voltage basis consisting of neighboring voltage drops. These choices are arbitrary, but are physically motivated and convenient. Explicitly, the  current basis $\mathcal{B}_I$ and voltage basis $\mathcal{B}_V$ read
\bea
\mathcal{B}_{I} &= \left(
\begin{array}{ccccccccc}
 1 & 0 & 0 & 0 & 0 & -1 & 0 & 0 & 1 \\
 0 & 1 & 0 & 0 & 0 & 0 & -1 & 0 & 1 \\
 0 & 0 & 1 & 0 & 0 & 0 & 0 & -1 & 1 \\
 0 & 0 & 0 & 1 & 0 & 0 & 0 & 0 & 1 \\
 -1 & 0 & 0 & 0 & 1 & 0 & 0 & 0 & 1 \\
 0 & -1 & 0 & 0 & 0 & 1 & 0 & 0 & 1 \\
 0 & 0 & -1 & 0 & 0 & 0 & 1 & 0 & 1 \\
 0 & 0 & 0 & -1 & 0 & 0 & 0 & 1 & 1 \\
 0 & 0 & 0 & 0 & -1 & 0 & 0 & 0 & 1 \\
\end{array}
\right), \\
\mathcal{B}_V &= \left(
\begin{array}{ccccccccc}
 1 & 0 & 0 & 0 & 0 & 0 & 0 & 0 & 1 \\
 -1 & 1 & 0 & 0 & 0 & 0 & 0 & 0 & 1 \\
 0 & -1 & 1 & 0 & 0 & 0 & 0 & 0 & 1 \\
 0 & 0 & -1 & 1 & 0 & 0 & 0 & 0 & 1 \\
 0 & 0 & 0 & -1 & 1 & 0 & 0 & 0 & 1 \\
 0 & 0 & 0 & 0 & -1 & 1 & 0 & 0 & 1 \\
 0 & 0 & 0 & 0 & 0 & -1 & 1 & 0 & 1 \\
 0 & 0 & 0 & 0 & 0 & 0 & -1 & 1 & 1 \\
 0 & 0 & 0 & 0 & 0 & 0 & 0 & -1 & 1 \\
\end{array}
\right) \ .
\eea
Note that both bases include $\mu$ as their last vector. The first $n-1$ column vectors in each basis matrix correspond to four (or three) terminal measurements of the resistance, and the last column is $\mu$, the vector spanning the nullspace. Let us define an $n\times n$ matrix $\mathcal{R}$ corresponding to the measured resistances where 
\bea
\mathcal{R}_{ij} &= \begin{cases}
R_{[\mathcal{B}_I]_j,[\mathcal{B}_V]_i} \frac{h}{2 C_V e^2}, & i,j \in \{1,\dots, N-1\} \\
0, & i\text{ or }j = N \\ 
\end{cases}, 
\eea
in other words, $\mathcal{R}_{ij}$ is the non-dimensionalized measured resistance corresponding to the current configuration in the $j$th column of $\mathcal{B}_I$ and the voltage configuration of the $i$th column of $\mathcal{B}_V$ for $i,j = 1,\dots N-1$, and $\mathcal{R}_{ij}$ is equal to zero in the last column and last row. In terms of the unknown conductance matrix $S$, this yields
\bea
\mathcal{B}_V^T S^+ \mathcal{B}_I &= \mathcal{R}
\eea
and thus we can determine $S^+$ from the known resistances via
\bea
\label{eq:finalLB}
S^+ &= {\mathcal{B}_V^{-1}}^T \mathcal{R} \mathcal{B}_I^{-1} \ . 
\eea
Recalling that  ${S^+}^+ = S$, we find that the conductance matrix $S$ is fully determined by the experimental measurements by applying the pseduo-inverse
\bea
\label{eq:final}
S = \Big( {\mathcal{B}_V^{-1}}^T \mathcal{R} \mathcal{B}_I^{-1} \Big)^+.
\eea
Recalling from Eq. \ref{eq:Smat} that $S_{ij} = \delta_{ij} \sum_k T_{ik} - T_{ij}$, we see that once $S$ is known, the off-diagonal elements of $T$ are known:
\bea
T_{ij} = - S_{ij}, \qquad \text{for } i \neq j \ . \\
\eea
The off-diagonals of $T$ are sufficient because we showed in Supplementary Information section \ref{app:gicc} that the diagonal elements are not physically meaningful. For ease, we simply set the full diagonal equal to zero. It is convenient at this stage to impose TRS by symmetrizing the $T$ matrix. We expect TRS to be preserved in our TDBG system, but due to noise in the data, the $T$ matrix inverted from the measurements is not guaranteed to be symmetric. This is remedied by defining $T^{(TRS)}_{ij} = 1/2 T_{ij} + 1/2 T_{ji}$. We compare the TRS-breaking $T$ matrix with $T^{TRS}$ in Fig. \ref{fig:constcheck} and find that there is little difference in the gapped regions. Henceforth, we will drop the TRS label and simply refer to the symmetrized matrix as $T$.

\subsection{Consistency Checks}

\begin{figure*}
 \centering
 \textbf{a} \includegraphics[width=7cm]{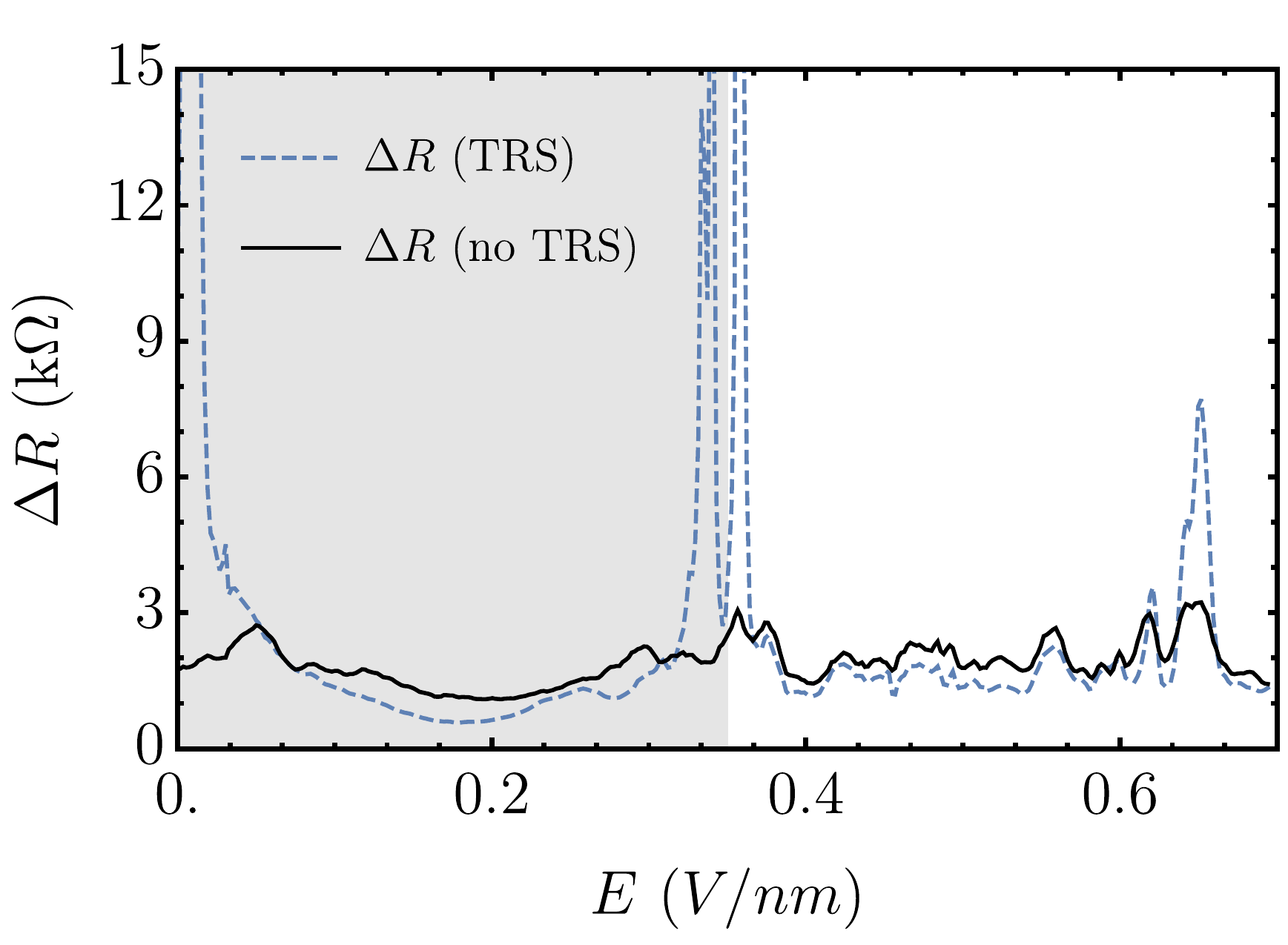}
 \textbf{b} \includegraphics[width=7cm]{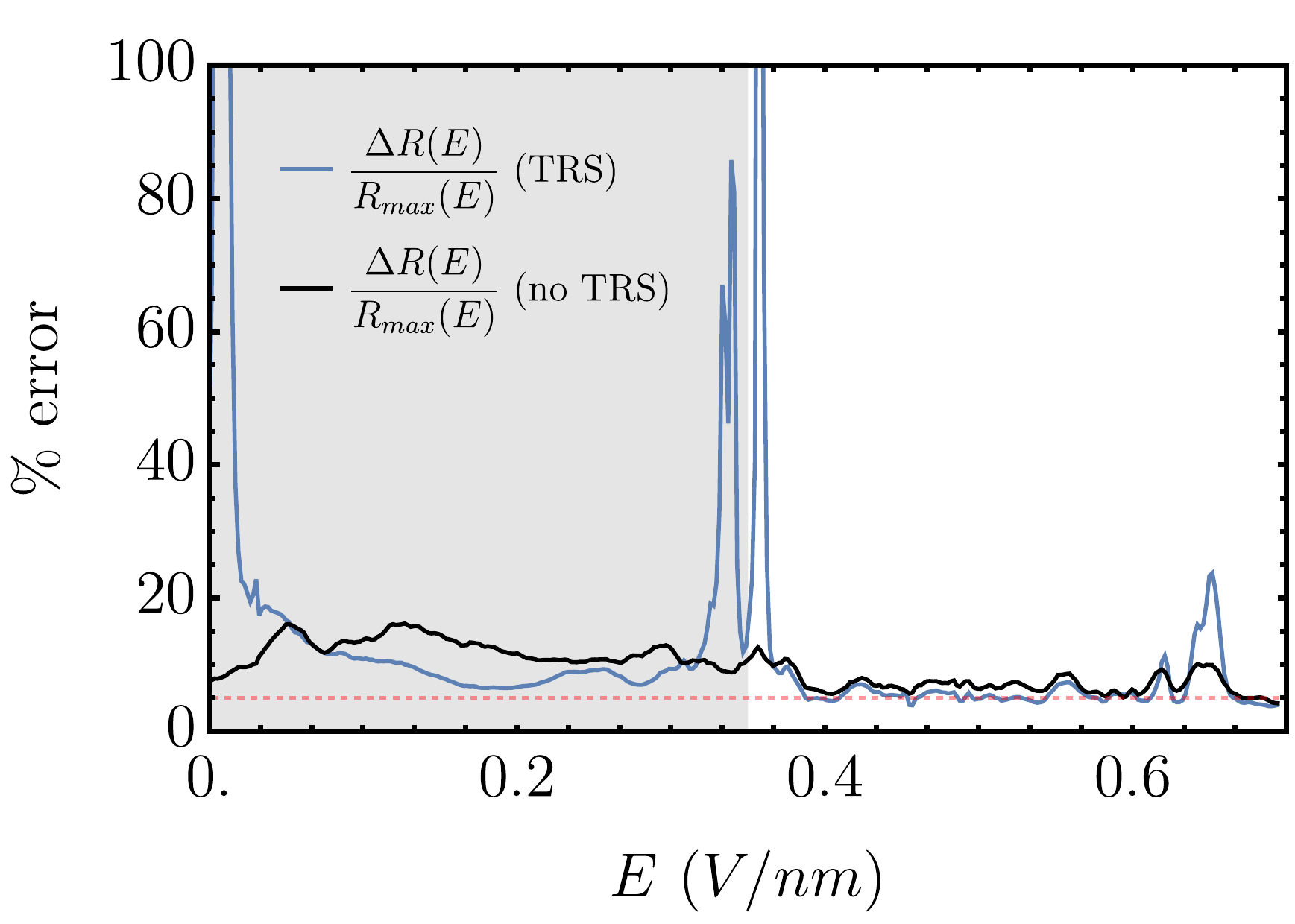}
\caption{\textbf{a}, We compare the resistance errors between the predicted $R$ values from the $T$ matrix and their measured counterparts. Away from the gap closing transition which divides the shaded and unshaded regions, $T$ and $T^{(TRS)}$ perform similarly well. \textbf{b}, We plot the resistance errors normalized by the largest measured resistance among the 43 data points at each $E$, and report the ratio as a percent error. The red dashed line highlights $5\%$ error.}
\label{fig:constcheck}
\end{figure*}

We have shown that $(n-1)^2$ resistance measurements are sufficient to exactly determine the observable (off-diagonal) entries of $T_{ij}$ via the formula in Eq. \ref{eq:finalLB}. This formula makes no assumptions on the form of $T$ so it is an exact description of the data, e.g. it is not a best-fit model. In our nine-terminal sample, we performed the 64 resistance measurements at each $E$ to determine $T_{ij}(E)$. Then, to check the consistency of the LB equation, we took 43 subsequent measurements (abbreviated $R^{exp}$) with different current/voltage configurations and compared the resistances to those predicted from $T$ (abbreviated $R^{predicted}$). We compute the error 
\bea
\Delta R = \sqrt{\frac{1}{43} \sum_{m=1}^{43} (R_m^{exp} - R_m^{predicted})^2} 
\eea
at each $E$ and plot the results in Fig. \ref{fig:constcheck}a. We find that, almost everywhere, the TRS and no TRS $T$ matrices perform similarly, achieving an error of $\Delta R \approx \SI{3} {k\Omega}$. The spikes in the TRS error occur exactly at the bulk gap closings where the bulk conductivity becomes nonzero. Note that $\Delta R$ takes into account all measured resistances, which range between $0$ and $\SI{30} {k\Omega}$, at a given $E$. We define the percent error at a given $E$ as $\Delta R/R_{max}$. This translates to an error of between $5\%$ and $10\%$ (see Fig. \ref{fig:constcheck}b). This fairly high accuracy confirms the validity of the LB equation and is an important consistency check on our formalism. 

\subsection{Topological Edge States}
\label{app:sigma}

One advantage of our exact solution of the LB equation is that it makes no assumptions on the form of the $T$ matrix, and hence our experimentally determined $T$ matrix is an exact representation of the data. Furthermore, $T$ performs very well when predicting the values of other resistances. For these reasons, the full $T$ matrix serves as an accurate and complete description of the edge physics of the TDBG sample. 

We would now like to use our $T$ matrix to show that the edge states of gapped TDBG at large enough $E$ have the properties we associate with topological phases. Firstly, we expect an insensitivity to impurities, defects of the edge, or more generally any change of parameter that does not change the bulk mirror Chern number. In our experiment, the transverse electric field $E$ is an excellent candidate. Once the gap is opened at $E = \SI{.33}{V/nm}$, increasing the field further does not change the Chern number although it does have a strong effect on the band structure. A signature of the topological origin of the edge states is their insensitivity to $E$ in the gapped regime. A simple but powerful way to show this is to study how the whole matrix $T(E)$ changes with $E$. We consider the following quantity
\bea
|| \Delta T(E) || \equiv  \frac{4 e^2}{h} \times \frac{1}{N(N-1)}\sqrt{ \sum_{i,j =1}^N (T_{ij}(E) - T_{ij}(\SI{0.7}{V/nm}))^2},
\eea
which compares $T(E)$ to $T(\SI{0.7}{V/nm})$ by averaging the squared difference of every entry of the two matrices. This is related to the Frobenius norm, but divided by the number of off-diagonal entries $N(N-1)$ so $||\Delta T(E)||$ does not scale with the number of terminals. We chose $\SI{0.7}{V/nm}$ as the $E$ field for comparison because it is the furthest into the gapped phase. We plot $||\Delta T||$ as a function of $E$ in Fig. 4f of the Main Text. We find that once the gap is open ($> \SI{.33}{V/nm}$), $\Delta T$ decays quickly and remains close to zero. This shows that the entire matrix, which fully characterizes the edge states, is essentially constant in the gapped phase and is thus insensitive to the transverse $E$ field. We contrast this case with non-topological edge states, like those on the zigzag edge of Bernal bilayer graphene which are gapped out by edge disorder or very large $E$ fields. We find the opposite behavior in our samples, which suggests nontrivial topology. Another strong comparison is to TDBG at $E < 0.33$V/nm where the system is a bulk conductor. We see strong variation in $||\Delta T||$, signaling high sensitivity to $E$. This contrasts the topological insulator phase.

We now discuss another measure of the edge states to assess the topology which generalizes the total conduction ($\sigma$) discussed by B\"uttiker\autocite{PhysRevB.38.9375,PhysRevLett.57.1761}. We define this order parameter (as in Eq. \ref{eq:sigma}) by
\begin{equation}
\sigma(E) = \frac{2 C_v e^2}{h} \times \frac{1}{2N}\sum_{i\neq j}^N T_{ij}(E)
\end{equation}
which is a sum over all the off-diagonal elements of $T$. Intuitively, this corresponds to the total edge-to-edge transmission. In a topologically trivial insulator, edge states --- if they exist --- are not protected, and $\sigma$ will be a small, non-quantized number which is sensitive to disorder, applied fields, and edge perturbations. In Fig. \ref{fig4}, we see that $\sigma \approx e^2/h$ in the topological phase. This robust, nonzero value provides plausiblility of the survival of topological edge states, but fewer than would be expected from the bulk valley Chern number $C_V =4$ (including spin). Hence, $\sigma$ quantitatively diagnoses the topology of the edge and shows that the bulk-boundary correspondence remains partially intact despite the edge symmetry-breaking. Another striking feature of Fig. 4f is the divergence in $\sigma(E)$ as $E \to 0.33$V/nm where the gap closes. Because $\sigma$ is a 1D (edge) conductivity, the closing of the gap creates a finite 2D (bulk) conductivity which sends $\sigma \to \infty$. Mathematically, this arises in our formalism due to the pseudo-inverse of the resistance matrix. A bulk conductor reduces the resistance, leading to some singular values of $\mathcal{R}$ approaching zero. This creates a divergence in the pseudo-inverse.

In topological insulators, we expect $\sigma$ to have a robust, nonzero value. To illustrate this, let us consider the case of a perfectly clean QSH insulator with mirror Chern number $C_M = (C_{\uparrow} - C_{\downarrow})/2 = 1$, meaning that there is one chiral edge state for each spin, and they propagate with opposite chiralities. These states are topologically protected from scattering, so weak disorder will not affect them. This system is described by
\bea
\label{eq:TQSHapp}
T^{QSH}_{ij} = \delta_{i,j+1} +\delta_{i+1,j} 
\eea
which describes perfect transmission between neighboring terminals in both directions. Note that $i,j$ should be understood mod $N$. Concretely, for $N =5$, $T$ takes the form
\bea
T^{QSH} &= \left(
\begin{array}{ccccc}
 0 & 1 & 0 & 0 & 1 \\
 1 & 0 & 1 & 0 & 0 \\
 0 & 1 & 0 & 1 & 0 \\
 0 & 0 & 1 & 0 & 1 \\
 1 & 0 & 0 & 1 & 0 \\
\end{array}
\right) \ .
\eea
We emphasize that because there are two edge states ($C_M = 1$ corresponds to a pair of counter-propagating edge states, e.g. one chiral state and one anti-chiral state), the columns of $T^{QSH}$ sum to 2. Hence for a probabilistic interpretation of $T$ as a scattering matrix, we must divide it by $2C_M$. To compute $\sigma$, we use the fact that all column sums are equal to 2, so
\bea
\sigma^{(QSH)} = \frac{C_M e^2}{h} \times \frac{1}{2N}\sum_{i\neq j}^N T^{(QSH)}_{ij} = C_M \frac{ e^2}{h}
\eea
which explains the $1/(2N)$ normalization we have chosen in Eq. \ref{eq:sigma}. Note that in this case, $\sigma$ is quantized in multiples of the conductance quantum. In a more realistic sample, there may be weak transmission across more distant terminals, but the columns sums of $T$ will still be equal to the number of edge states, up to uncertainties in measurements. (Column sums of less than the number of edge states implies scattering.) Similarly, topological edge states are protected from weak disorder and back-scattering. Thus, although $T$ is generically less structured than Eq. \ref{eq:TQSHapp}, $\sigma$ is still expected to be fairly well quantized. 

However, in TDBG, the situation is more complicated because the valley symmetry protecting the topological index is expected to be broken on the edge because it arises from the geometric moir\'e pattern. This differs from the QSH model where spin conservation protects the mirror Chern number and is not broken on the edge. As such, some back-scattering is expected, and indeed it is \textit{a priori} not at all clear that any edge states would survive. However, we find in our experiments that $\sigma(E)$ reaches a finite, nonzero value in the gapped phase of approximately one quarter of the $4e^2/h$ conductance we would expect from a $C_V=2$ with a spin degeneracy, as shown in Fig. 4f of the Main Text.

Lastly, we will show that $\sigma$ is a natural observable in the LB formalism because it is an invariant under a discrete gauge symmetry corresponding to the arbitrary labeling of terminals $i=1,\dots,N$ which amounts to a permutation symmetry of the LB equation. Under a relabeling of the terminals $i \to \pi_i$ where $\pi$ is a permutation of the numbers $1,\dots,N$, there is a corresponding transformation of the conductance matrix: $S \to P S P^T$ where $P_{ij} = \delta_{i,\pi_j}$, and similarly $T \to P T P^T$. We remark that permutation transforms of this type simply permute the diagonal elements of $T$ among themselves. This is consistent with all diagonal elements being unobservable. Because the physics of the edge is invariant under this relabeling, any well-defined order parameter of $T$ must be invariant under $T \to P T P^T$. We now show that $\sigma$ satisfies this requirement. One simple way is to rewrite $\sigma$ using the vector $\mu$ with all entries equal to $1$. We find
\bea
\label{eq:sigmainv}
\sigma = \frac{2 C_v e^2}{h} \times \frac{1}{2N} \left( \mu^T T \mu - \text{Tr } T \right) \ .
\eea
The trace term is invariant under $T \to P T P^T$ using the cyclicity property, and the fact that $P$ is orthogonal. The first term $\mu^T T \mu$  is invariant because $P \mu = \mu$, i.e. $\mu$ is an eigenvector of all permutations matrices because all its entries are the same. Note that the diagonal entries of $T$ cancel from Eq. \ref{eq:sigmainv}.

\section{A Comparative Analysis of Bernal Bilayer Graphene}
\label{app:bernal}

In this Appendix, we briefly introduce Bernal bilayer graphene\autocite{2011SSCom.151.1075P,2016PhR...648....1R,PhysRevB.92.115437} from the perspective of valleytronics\autocite{2016NatRM...116055S}. We emphasize that there is no bulk valley symmetry in Bernal graphene, and hence there is not a bulk valley Chern number to protect edge states. Hence it is important to distinguish TDBG as a topologically protected state, whereas Bernal graphene is not. Indeed, Fig. 3b of the Main Text shows a monotonic increase in the longitudinal resistance as a function of $E$, indicating a trivial insulator. 

However, Bernal graphene does have strongly peaked Berry curvature with opposite sign in the $K$ and $K'$ valleys. The cancellation of the Berry curvature causes the global Chern number to vanish, but there are still non-topological edge states that appear because the Berry curvature in each valley is large. For an armchair termination on cylinder boundary conditions, no edge states appear because the valleys are projected onto each other and cancel, but on zigzag terminations which do not project the valleys onto each other, edge states do appear in the spectrum\autocite{2011SSCom.151.1075P}. However, these edge states are not topological and can be removed by edge perturbations. 

We can also understand the local lack of stability of the Bernal edge states by calculating their characteristic localization length $\xi$. Recall that the low energy behavior of Bernal graphene is a quadratic band touching\autocite{2016PhR...648....1R} with a Berry phase of $2\pi$. When applied electric fields open a gap of $V \sim \SI{.1} {eV}$, as is reachable by experiment, we can estimate the spread of the Berry curvature from the quadratic band touching using $\hbar^2 v_F^2 k^2/t = V$ where $v_F = (\SI{610} {meV \cdot nm})/\hbar$ is the Fermi velocity of graphene and $t = \SI{3.1} {eV}$ is the hopping parameter. This gives an estimate of the characteristic momentum scale $k \approx \SI{.9} {nm^{-1}}$, or a real space length of $\xi \approx \SI{1} {nm}$. Edge roughness on this order will gap the non-topological edge states. 

\begin{figure*}
 \centering
 \includegraphics[height=5.5cm]{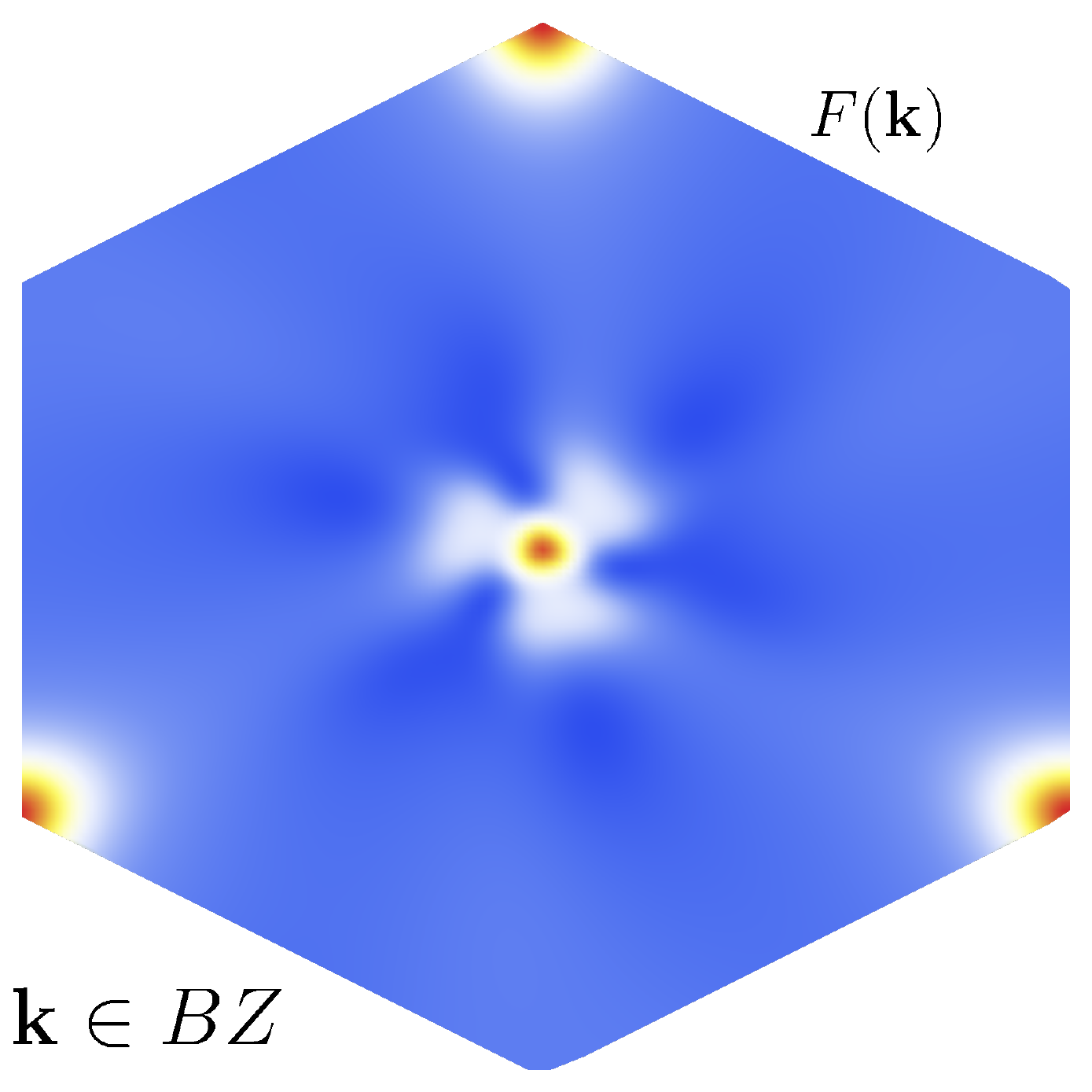}\includegraphics[height=5.5cm]{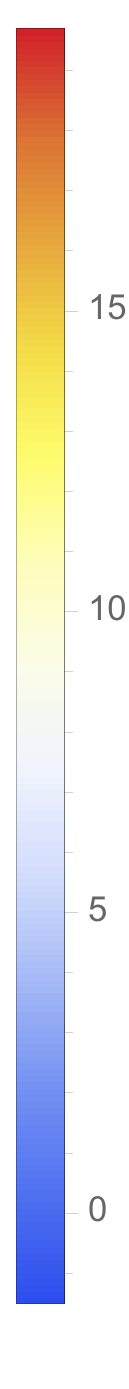}
 \caption{We show the Berry curvature $F(\mbf{k})$ normalized such that $\sum_{\mbf{k}} F(\mbf{k}) = 2$, the Chern number of the TDBG Hamiltonian in the $K$ valley. We see strong peaks in $F(\mbf{k})$ at the $\Gamma_M$ and $K_M$ points which are of the same sign. The area of these peaks (defined by $F(\mbf{k}) > 10$) is approximately $1.4\%$ of the moir\'e Brillouin zone, leading to a characteristic momentum scale of  $0.1 k_M$. }
\label{fig:TDBGF}
\end{figure*}

In comparison, we now calculate the Berry curvature $F(\mbf{k})$ of TDBG in the topological phase at $\theta = 1.60^\circ$ using Eq. \ref{eq:chern} (see Fig. \ref{fig:TDBGF}). We find that the characteristic momentum scale of the Berry curvature is $k \sim 0.1 k_M$, giving a real space length scale of approximately 10 moir\'e periods. At $\theta = 1.60^\circ$, this is approximately $\SI{100} {nm}$. The increased delocalization of the edge states in TDBG explains their survival in our samples, despite the breakdown of the valley symmetry on the edge. Because the edge states are extended over many moir\'e periods, they are less sensitive to the edge termination, and retain the global symmetry-enforced topological protection of the bulk valley Chern number. Our calculations of $\xi$ agree at orders of magnitude with other heuristics of edge states localization, such as band flatness: $\xi/a = \sqrt{ \frac{W}{\Delta}}$ where $W$ is the bandwidth, $\Delta$ is the band gap, and $a$ is the characteristic length scale. For Bernal stacked graphene, $a = \SI{2.46} {\angstrom}$, $W \sim 1$ eV, and $\Delta \sim 10-100$ meV lead to $\xi \sim 1$ nm. 

\end{document}